\documentclass[%
 reprint,
 superscriptaddress,
 amsmath,amssymb,
 aps,
 prb,
]{revtex4-2}

\setcitestyle{number}
\usepackage{graphicx}
\usepackage{amsmath,amsfonts,amssymb}
\usepackage{dcolumn}
\usepackage{bm}
\usepackage{hyperref}
\hypersetup{
    colorlinks=true,
    linkcolor=blue,
    filecolor=magenta,      
    urlcolor=blue,
    citecolor=blue
    }

\usepackage{blindtext}

\begin{document}


\title{Variable Temperature and Carrier-Resolved Photo-Hall Measurements of High-Performance Selenium Thin-Film Solar Cells}

\author{Rasmus S. Nielsen}
\email[]{Electronic mail: rasmus.nielsen@empa.ch}
\affiliation{Transport at Nanoscale Interfaces Laboratory, Swiss Federal Laboratories for Material Science and Technology (EMPA), Ueberlandstrasse 129, 8600 Duebendorf, Switzerland}

\author{Oki Gunawan}
\affiliation{IBM Thomas J. Watson Research Center, 1101 Kitchawan Road, Yorktown Heights, NY 10598, USA}

\author{Teodor Todorov}
\affiliation{IBM Thomas J. Watson Research Center, 1101 Kitchawan Road, Yorktown Heights, NY 10598, USA}

\author{Clara B. Møller}
\affiliation{SurfCat, DTU Physics, Technical University of Denmark, 2800 Kongens Lyngby, Denmark}

\author{Ole Hansen}
\affiliation{National Center for Nano Fabrication and Characterization (DTU Nanolab), Technical University of Denmark, 2800 Kongens Lyngby, Denmark}

\author{Peter C. K. Vesborg}
\affiliation{SurfCat, DTU Physics, Technical University of Denmark, 2800 Kongens Lyngby, Denmark}

\begin{abstract}
Selenium is an elemental semiconductor with a wide bandgap suitable for a range of optoelectronic and solar energy conversion technologies. However, developing such applications requires an in-depth understanding of the fundamental material properties. Here, we study the properties of the majority and minority charge carriers in selenium using a recently developed carrier-resolved photo-Hall technique, which enables simultaneous mapping of the mobilities and concentrations of both carriers under varying light intensities. Additionally, we perform temperature-dependent Hall measurements to extract information about the acceptor level and ionization efficiency. Our findings are compared to results from other advanced characterization techniques, and the inconsistencies are outlined. Finally, we characterize a high-performance selenium thin-film solar cell and perform device simulations to systematically address each discrepancy and accurately reproduce experimental current-voltage and external quantum efficiency measurements. These results contribute to a deeper understanding of the optoelectronic properties and carrier dynamics in selenium, which may guide future improvements and facilitate the development of higher-efficiency selenium solar cells.
\end{abstract}


\maketitle


\section{Introduction}

Selenium has reemerged as a promising inorganic semiconductor with desirable properties for a range of optoelectronic applications \cite{todorov2017a, zhu2019a, youngman2021a}. In its trigonal phase, selenium features a direct bandgap in the range of 1.8 to 2.0 eV, tunable through alloying with tellurium \cite{hadar2019b, zheng2022a, deshmukh2022a}. Combined with its high absorption coefficient ($\alpha>10^5$ cm$^{-1}$) in the visible spectrum \cite{nielsen2022a}, and long-term air stability \cite{liu2020a, zhu2016a}, selenium could be an ideal candidate for use in photodiodes \cite{li2024a, chen2024a, adachi2023a}, indoor solar cells \cite{yan2022a, wang2024a, fujimura2022a, wei2023a, ieee2017a}, and, as recently demonstrated, as the top cell in tandem solar cells \cite{nielsen2024a}. However, despite the integration into a range of applications, and a recent advance in the record device efficiency \cite{lu2024a}, a high-efficiency photovoltaic device using selenium has yet to be realized, questioning its potential as the optoelectronic material of tomorrow.

In the design and engineering of high-performance optoelectronic devices, understanding and controlling material properties, particularly carrier properties, is crucial \cite{crovetto2024a}. In solar cells, it is the majority carriers that largely dictate the overall device architecture, the width of the depletion region and the bulk series resistance. Conversely, the properties of the minority carriers determine key performance parameters, such as the recombination lifetime, diffusion length, and recombination coefficients, all of which directly impact the efficiency of the device \cite{kirchartz2018a}. However, characterizing the properties of both majority and minority carriers requires a broad range of experimental techniques, often involving different sample configurations and varying conditions, such as steady-state versus dynamic responses under different illumination levels and temperature \cite{gokmen2013a}. This diversity in measurement conditions adds complexity to the analysis and interpretation of the carrier properties, potentially leading to inconsistencies reported in the literature. For selenium, a general lack of thorough experimentation has further contributed to the unclear understanding of its carrier properties, leaving the material's true photovoltaic performance potential somewhat ambiguous. 

\begin{figure*}[t!]
    \centering
    \includegraphics[width=0.8\textwidth,trim={0 0 0 0},clip]{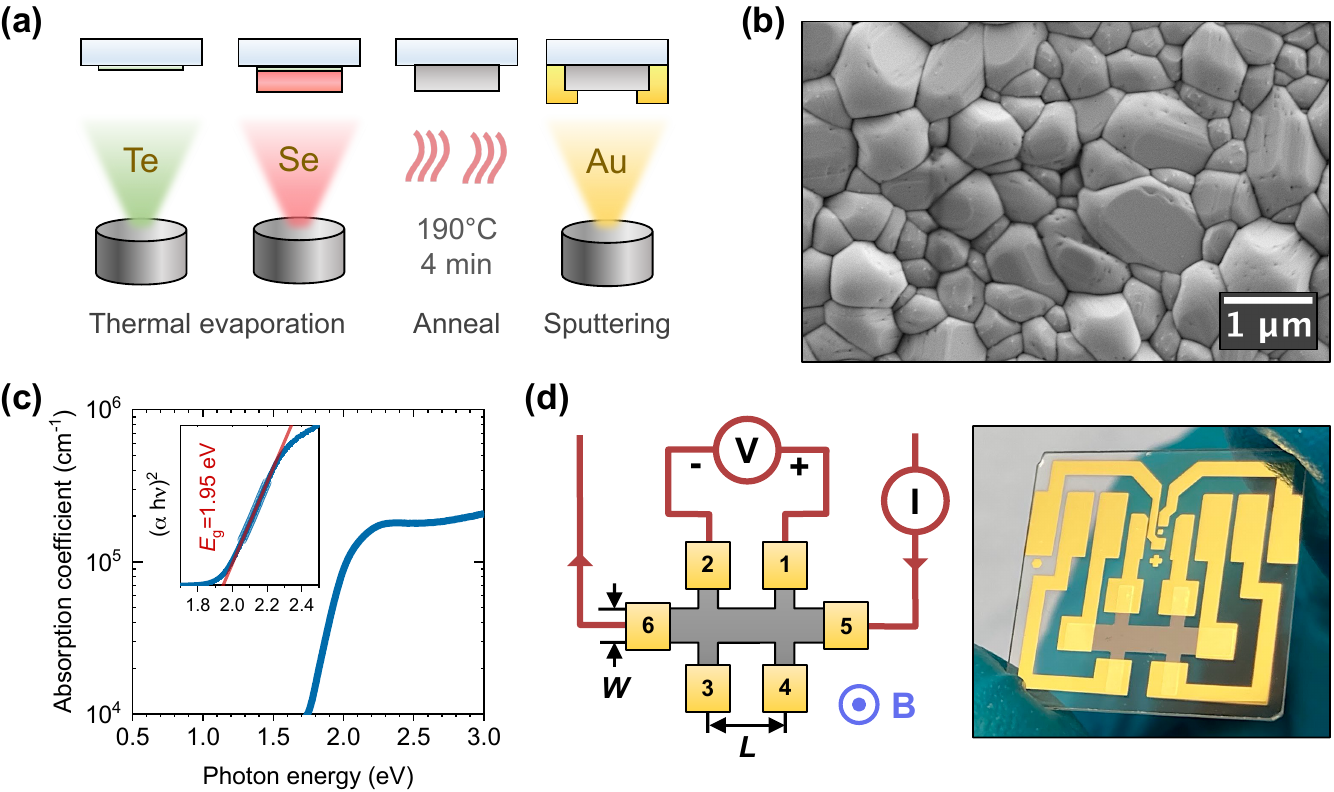}
    \caption{Fabrication and characterization of the selenium Hall bar on quartz. (a) Schematic of the thermal evaporation, annealing, and sputtering processes. (b) Top-view SEM image of the polycrystalline selenium thin-film. (c) Absorption coefficient of the selenium thin-film obtained from UV-vis spectroscopy, along with the corresponding Tauc plot used to extrapolate the optical direct bandgap. (d) Schematic representation and photograph of the six-terminal selenium Hall bar.}
    \label{fig:Figure1}
\end{figure*}

In this work, we used the recently developed carrier-resolved photo-Hall technique to simultaneously study the properties of both charge carriers as a function of light intensity \cite{gunawan2019a}. These results were complemented by temperature-dependent Hall measurements to gain additional insight into the influence of non-idealities in the selenium thin-film. By extending our model to account for carrier freeze-out and depletion effects at surfaces and interfaces, we arrived at a more nuanced interpretation of the carrier dynamics in selenium. We further validated this extended model by fabricating and characterizing state-of-the-art selenium solar cells in parallel with the selenium Hall bar, and incorporated the properties of the charge carriers determined from the photo-Hall measurements into device simulations, successfully reproducing experimental current-voltage characteristics and quantum efficiency measurements. Finally, we compared these results with our previous findings from transient THz spectroscopy, concluding that the photovoltaic performance potential of selenium, particularly due to much higher carrier mobilities, is greater than first anticipated. We also emphasize that addressing non-radiative recombination losses and improving carrier lifetimes through defect engineering is the most critical next step in developing higher-efficiency selenium solar cells.

\section*{Hall bar sample preparation}

The selenium Hall bar was fabricated on fused quartz substrates using an in-house process flow optimized for high-performance selenium thin-film solar cells. These photovoltaic devices have demonstrated record open-circuit voltages close to 1 V and power conversion efficiencies exceeding 5\% \cite{nielsen2022a}. The fabrication process involves thermal evaporation, annealing, and sputtering to deposit the selenium thin-film and gold contacts onto fused quartz substrates, as illustrated schematically in Fig \ref{fig:Figure1}(a). The resulting polycrystalline selenium thin-film exhibits large crystal grains, as shown in the top-view SEM image in Fig \ref{fig:Figure1}(b). To determine the absorbed photon density in the carrier-resolved photo-Hall measurements, the absorption coefficient of the selenium thin-film was measured using UV-vis spectroscopy, as shown in Fig \ref{fig:Figure1}(c). The corresponding Tauc plot, included as an inset, was used to extrapolate the optical direct bandgap.

The six-terminal Hall bar, with an active area of 2 mm x 4 mm and a film thickness of 0.30 $\mu$m, is designed to minimize the influence of voltage drops along the current-carrying path and cancel out spurious voltages, thus improving the accuracy of the Hall voltage measurements. This device geometry was realized using dedicated shadow masks during the physical vapor deposition steps, with the schematic and a photograph of the final device depicted in Fig \ref{fig:Figure1}(d). 

Additional experimental details are provided in the methods section.

\section{Carrier-resolved photo-Hall measurements}

\begin{figure*}[t!]
    \centering
    \includegraphics[width=0.8\textwidth,trim={0 0 0 0},clip]{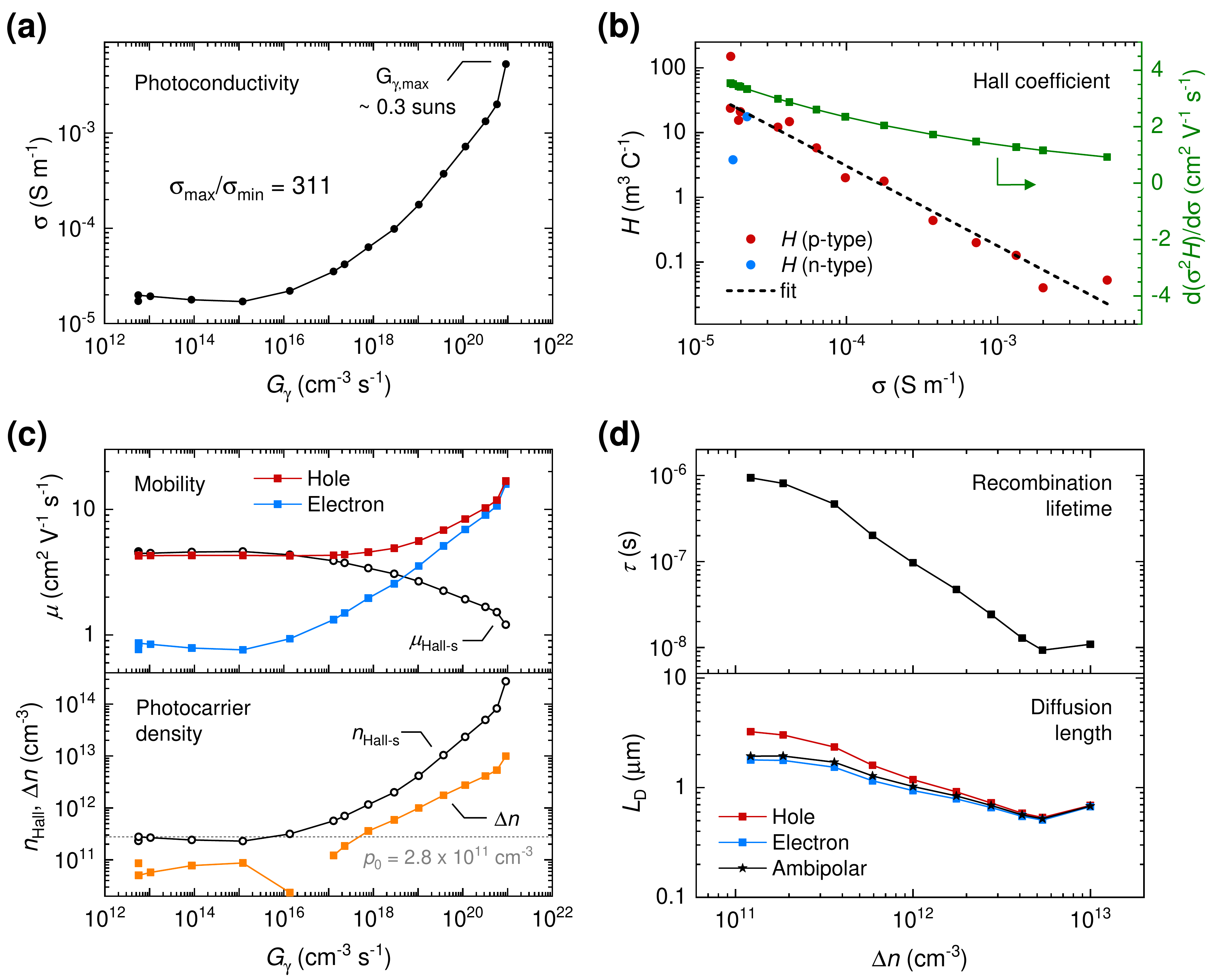}
    \caption{Carrier-resolved photo-Hall analysis of the polycrystalline selenium thin-film. (a) Photoconductivity $\sigma$ plotted against absorbed photon density $G_\gamma$. (b) Hall coefficient $H$ plotted against $\sigma$. The slope of the $\sigma$-$H$ curve contains detailed information about the mobilities of the majority and minority carriers, where specifically, the derivative d($\sigma^\text{2}H$)/d$\sigma$ corresponds to the Hall mobility difference $\Delta \mu_\text{Hall}$. (c) Majority and minority carrier mobilities and the photocarrier density $\Delta n$ plotted against $G_\gamma$. Here, $n_\text{Hall-s}$ and $\mu_\text{Hall-s}$ denotes the single-carrier Hall carrier density and mobility. The background carrier density $p_0$ is indicated by a grey line. (d) Recombination lifetime $\tau$ and diffusion length $L_\text{D}$ plotted against $\Delta n$.}
    \label{fig:Figure2}
\end{figure*}

First, we measured the sample in the dark to determine the properties of the majority carriers. These measurements revealed that the selenium Hall bar is p-type, with a dark carrier density of $p_0=\text{2.8}\times\text{10}^\text{11}$ cm$^\text{-3}$ and a Hall mobility of $\mu_p=\text{4.5}$ cm$^\text{2}$ V$^\text{-1}$ s$^\text{-1}$. Subsequently, we measured the sample under various illumination levels using a laser (wavelength $\lambda=\text{520}$ nm, intensity up to 31 mW cm$^\text{-2}$). The laser intensity was monitored using a photodiode, and the absorbed photon density $G_\gamma$ was calculated based on the absorption coefficient, transmission, and reflectivity of the selenium Hall bar. For each light intensity, longitudinal and transverse magnetoresistances were measured, and the corresponding photoconductivity $\sigma$ and Hall coefficient $H$ were calculated. These results are presented in Fig \ref{fig:Figure2}(a) and (b), showing that both $\sigma$ and $H$ change substantially upon illumination. Specifically, $\sigma$ increases by a factor of 311, while $H$ decreases by approximately three orders of magnitude.

In classical Hall measurements, the Hall mobility of the majority carriers is calculated as $\mu_\text{Hall}=\sigma H$. However, in photo-Hall measurements, the relationship between $\sigma$ and $H$ provides key insights into the Hall mobilities of both the majority and the minority carriers. In 2019, Gunawan et al. showed that the difference in Hall mobilities between the two carrier types, $\Delta \mu_\text{Hall}$, is related to the photoconductivity and Hall coefficient by the equation $\Delta\mu_\text{Hall}=\text{d}(\sigma^\text{2}H)/\text{d}\sigma$ \cite{gunawan2019a}. This relationship is plotted on the secondary y-axis in Fig \ref{fig:Figure2}(b), illustrating that the difference in Hall mobility between electrons and holes approaches zero as illumination increases. 

To calculate the carrier-resolved mobilities, we first need to solve the photo-Hall transport problem. An exact solution for p-type materials has been explicitly derived in Ref \cite{gunawan2019a}, providing analytical expressions for the mobility ratio $\beta=\mu_n/\mu_p$ and the photocarrier density $\Delta n$. Using these equations, we calculate the mobilities of holes ($\mu_p=\Delta \mu / \left( 1-\beta \right)$) and electrons ($\mu_n=\beta \mu_p$), as well as the excess carrier concentration. The results are presented in Fig \ref{fig:Figure2}(c), along with the single-carrier Hall mobility $\mu_\text{Hall-s}$ and carrier density $n_\text{Hall-s}$. Both $\mu_p$ and $\mu_n$ are observed to increase with $G_\gamma$, potentially due to a light-modulated intragranular barrier effect \cite{fowler1961a, dresner1964a}, the filling of electronic traps or defects and  increased screening at higher carrier densities. As expected, the photocarrier concentration also increases with $G_\gamma$.  Finally, we demonstrate that the single-carrier Hall analysis can yield values that differ significantly from the actual values of $\mu_p$, $\mu_n$, and $\Delta n$ obtained through the $\Delta \mu$ calculation model, underscoring the importance of specifying illumination conditions and the calculation models used when reporting these measurements. 


\begin{figure*}[t!]
    \centering
    \includegraphics[width=0.718\textwidth,trim={0 0 0 0},clip]{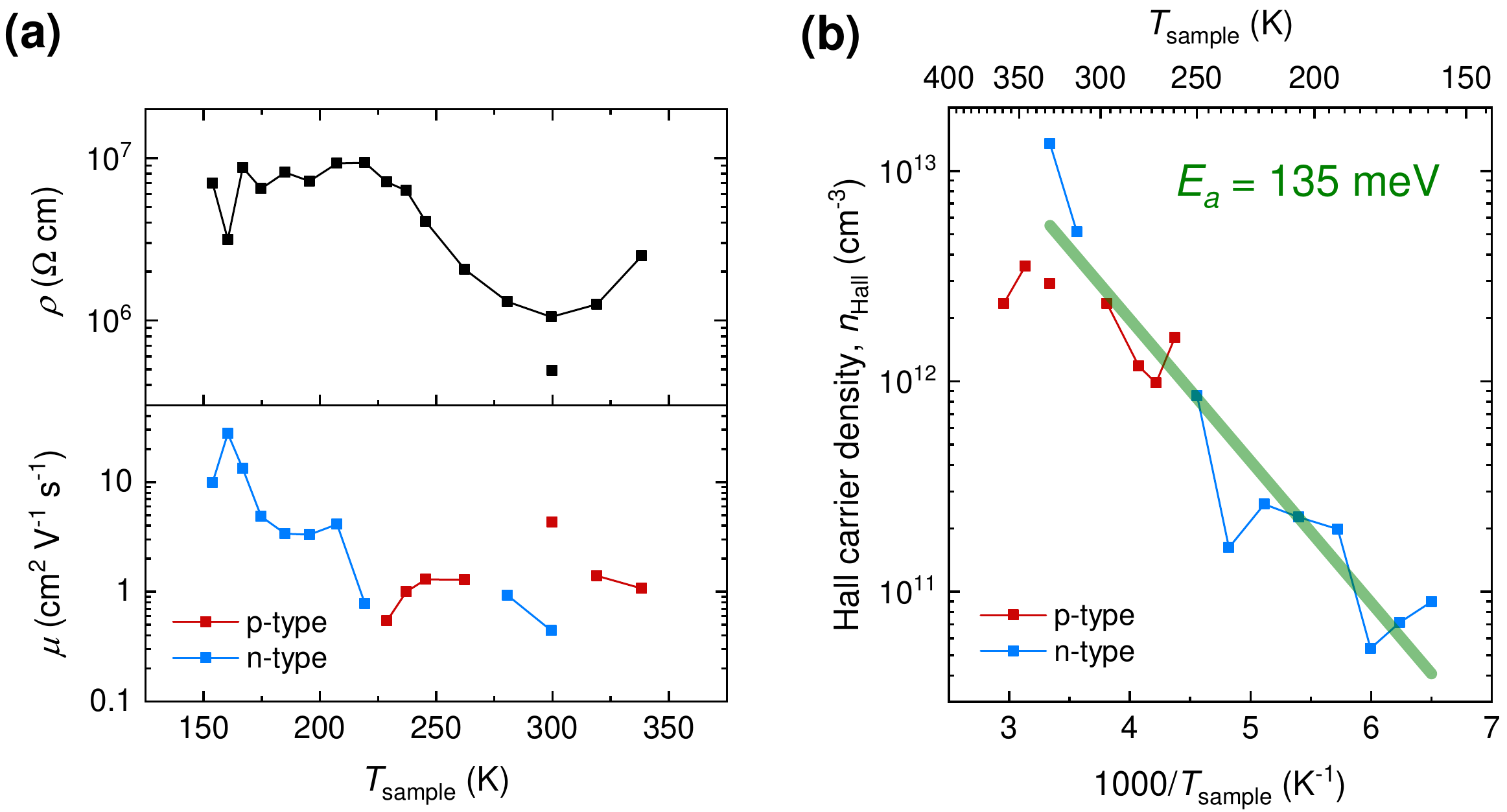}
    \caption{Temperature-dependent Hall analysis of the polycrystalline selenium thin-film in the dark. (a) Resistivity and Hall mobility as a function of sample temperature. (b) Arrhenius plot showing the natural logarithm of the Hall carrier density versus the reciprocal of the sample temperature. The apparent activation energy is fitted to $E_a = 135$ meV.}
    \label{fig:Figure3}
\end{figure*}

In addition the fundamental properties of the charge carriers, it is also possible to calculate derivative parameters such as the recombination lifetime, $\tau=\Delta n/G_\gamma$, and the carrier diffusion length, $L_\text{D}=\sqrt{k_\mathrm{B}T\mu\tau/q_e}$, where $k_\mathrm{B}$ is the Boltzmann constant and $T$ is the temperature. These parameters are plotted as a function of $\Delta n$ in Fig \ref{fig:Figure2}(d). The ambipolar diffusion length, $L_\text{D,am}=\sqrt{k_\mathrm{B}T\tau\left(n+p\right)/q_e\left(n/\mu_p+p/\mu_n\right)}$, has also been included, as it is more appropriate under high injection conditions ($\Delta n,\Delta p > p_0$); however, the difference between $L_\text{D}$ and $L_\text{D,am}$ is negligible. These results demonstrate that the recombination lifetime and carrier diffusion length are strongly dependent on the illumination conditions, further emphasizing the critical importance of specifying these conditions when reporting Hall measurements.

In summary, the carrier-resolved photo-Hall measurements showed that hole mobilities $\mu_p$ increase from 4.3 to 17 cm$^\text{2}$ V$^\text{-1}$ s$^\text{-1}$ and electron mobilities $\mu_n$ increase from 0.8 to 16 cm$^\text{2}$ V$^\text{-1}$ s$^\text{-1}$ with increasing illumination. The recombination lifetime $\tau$ reached 0.94 $\mu \text{s}$ with a corresponding carrier diffusion length of $L_\text{D}\approx2 \,\mu \text{m}$ at the lowest light intensity. At the highest light intensity, equivalent to 0.3 suns, the recombination lifetime decreased to 9.3 ns, and the carrier diffusion length to 0.5 µm. The observed trends strongly indicate that under 1 Sun conditions, the carrier mobilities will likely exceed 17 cm$^\text{2}$ V$^\text{-1}$ s$^\text{-1}$, while the recombination lifetime and diffusion length could decrease below 9.3 ns and 0.5 µm, respectively. It is important to note that these parameters vary substantially with photogenerated carrier density, as this light dependence is crucial for understanding and predicting photovoltaic performance potential. For instance, the excess carrier concentration at open-circuit conditions is generally higher than at the maximum power point, with the latter being the more relevant measure for predicting power conversion efficiencies.

\section{Temperature-dependent Hall measurements}

In addition to the carrier-resolved photo-Hall measurements, we also conducted temperature-dependent Hall measurements to study the mobility scattering mechanisms and the nature of the acceptor states in the selenium thin-film. However, due to the inherently low carrier concentration in the selenium thin-film, even at room temperature, minor fluctuations can significantly impact the Hall measurements, leading to noisy data and occasional flips in the sign of the Hall coefficient. This phenomenon is evident in Fig. \ref{fig:Figure3}(a), where the resistivity approaches $\text{10}^\text{7} \, \Omega \, \text{cm}$, and the majority carrier type appears to oscillate between p-type and n-type. This should not be interpreted as actual changes in the polarity of the selenium thin-film but rather as artifacts resulting from the extremely low Hall density and the high resistivity of the sample in the dark. Despite these challenges, the Arrhenius plot of the Hall carrier density shown in Fig. \ref{fig:Figure3}(b) reveals that the carrier density changes by two orders of magnitude across the investigated temperature range. Fitting this data to the Arrhenius equation yields an apparent activation energy of $E_a$ = 135 meV, indicating significant carrier freeze-out at room temperature.

\begin{figure*}[t!]
    \centering
    \includegraphics[width=0.85\textwidth,trim={0 0 0 0},clip]{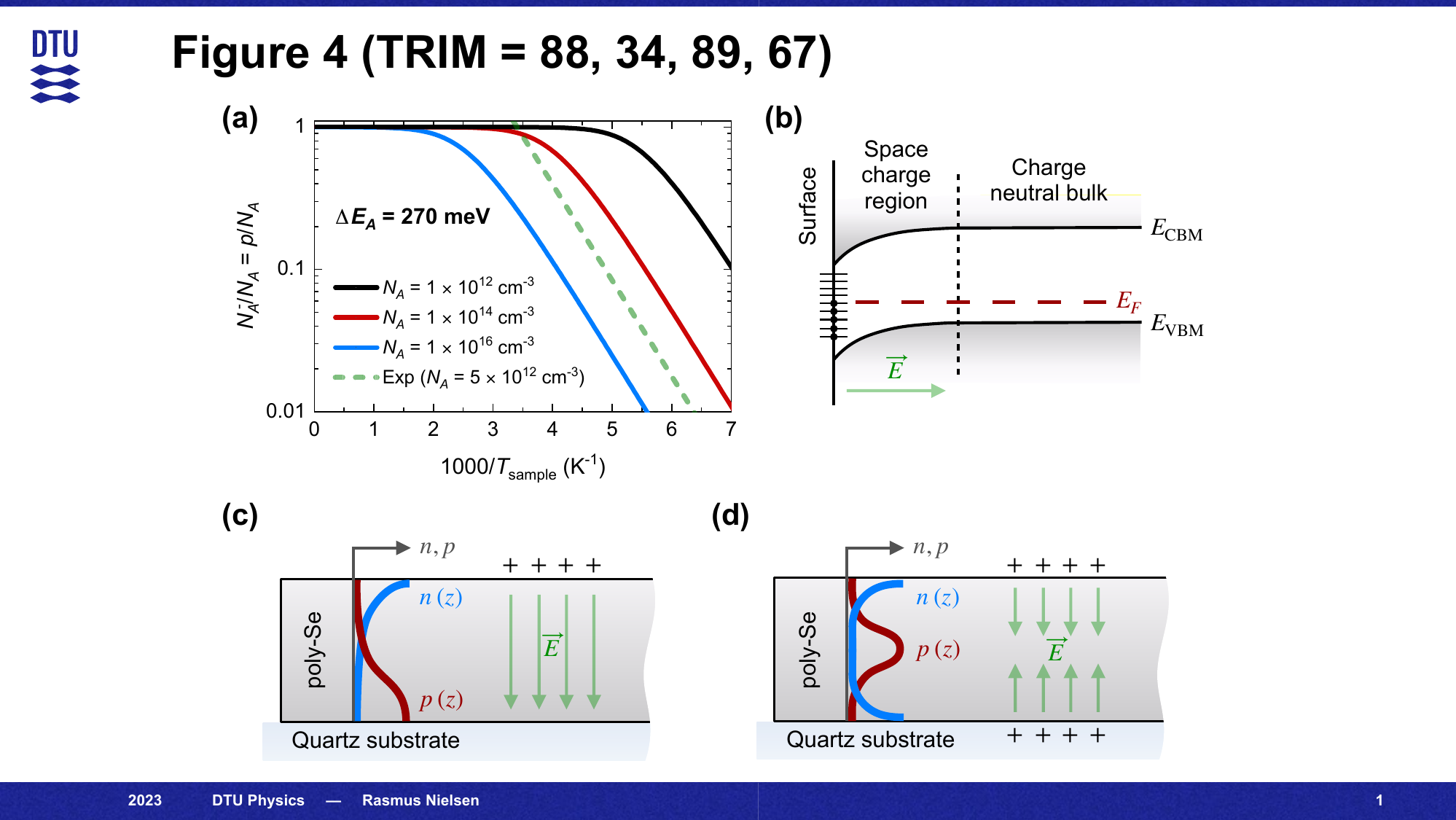}
    \caption{Possible origins of a discrepancy between the carrier and doping densities. (a) Arrhenius plot comparing the theoretical ionization efficiency of an acceptor with an ionization energy of $\Delta E_A$ = 270 meV at various concentrations (\textit{solid lines}) to the experimentally measured carrier density (\textit{dashed line}), highlighting the influence of carrier freeze-out. (b) Schematic of the energy band diagram, showing carrier depletion near the surface due to defects and Fermi-level pinning. (c) Spatial distribution of charge carriers affected by surface defects. (d) Expanded schematic illustrating the combined effects of surface and interface defects on the carrier distributions.}
    \label{fig:Figure4}
\end{figure*}

The activation energy obtained from the Arrhenius plot can be interpreted in several ways. One interpretation is that the activation energy corresponds to half of the ionization energy of an acceptor in the selenium film, implying that the acceptors require an energy of approximately $\Delta E_A$ = 270 meV for full ionization \cite{blakemore1969a, seeger1991a}. Another possible interpretation involves the behavior of the Fermi level, which may be pinned at the surface of the selenium film. As the temperature changes, the position of the pinned Fermi level can shift, affecting the observed activation energy. This shift occurs because the energy level at which the Fermi level is pinned can move with temperature due to changes in the surface states or the interaction between the Fermi level and the band edges. Consequently, the apparent activation energy might reflect the movement of the Fermi level rather than the intrinsic properties of the acceptor. However, this scenario is quite complicated to analyze, as it depends on factors such as doping level, film thickness, density of surface states, and whether the energy level representing a charge-neutral surface is also influenced by changes in temperature. This highlights the complexity of analyzing temperature-dependent Hall data in highly resistive, low-carrier-density materials like selenium.



\section*{Carrier density vs doping density}

In Hall effect measurements conducted under dark conditions, it is often assumed that the carrier density at room temperature is equal to the doping density, i.e., $p_0\approx N_A$ for a p-type semiconductor. This approximation holds if all acceptors are fully ionized and there are no compensating impurities or defects that neutralize the doping. However, given the significant carrier freeze-out indicated by our temperature-dependent Hall measurements and literature reports suggesting doping densities on the order of $N_A \sim 10^{16}\,\text{cm}^{-3}$ \cite{nielsen2022a, nielsen2023b, chen2024a, fu2022a, zheng2022a, yan2022a}, it is important to examine the ionization efficiency of the acceptors and consider potential carrier depletion due to defects. These factors are crucial for accurately interpreting the results of our photo-Hall measurements.

\subsection{Carrier Freeze-out}

First, we consider the phenomenon of carrier freeze-out, where not all acceptors contribute to the free carrier population at room temperature. To model the influence of carrier freeze-out on the discrepancy between the doping density and the carrier density, we derive the acceptor ionization efficiency assuming charge neutrality and no donor compensation (see Supplementary Information P.1). The ionization efficiency is then given by:

\begin{equation*}
    \frac{p}{N_A} = \frac{N_A^-}{N_A} = \frac{\sqrt{1+4g_A \left[\frac{N_A}{N_V}\right] \exp{\left(\frac{\Delta E_A}{k_\mathrm{B}T}\right)}}-1}{2g_A \left[\frac{N_A}{N_V}\right] \exp{\left(\frac{\Delta E_A}{k_\mathrm{B}T}\right)}}
\end{equation*}

\noindent where $p$ is the hole density, $N_A$ is the acceptor density, $g_A$ is the degeneracy factor of the acceptor level, $N_V$ is the effective density of states in the valence band, $\Delta E_A$ is the ionization energy of the acceptor, $k_B$ is the Boltzmann constant, and $T$ is the temperature.

Figure \ref{fig:Figure4}(a) shows an Arrhenius plot of the ionization efficiency as a function of temperature for various acceptor densities, assuming $g_A = 4$ and $N_V=1.64\times10^{20}\,\text{cm}^{-3}$ \cite{nielsen2022a}. The slope corresponding to an ionization energy of $\Delta E_A = 270$ meV aligns well with the experimental data. However, the shift of the curve suggests that even if all acceptors were fully ionized at room temperature, both the acceptor and carrier density should be orders of magnitude higher ($p=N_A > 10^{14}$ cm$^{-3}$). This observation implies that while partial freeze-out of the acceptors may be occurring in the selenium thin-film, the freeze-out model alone cannot account for the low carrier densities measured at room temperature. Since the absolute carrier densities observed are significantly lower than those predicted by the freeze-out model, additional mechanisms, such as surface and interface-related depletion effects, are likely contributing to the reduced carrier density in the selenium thin-film.

\subsection{Carrier Depletion at Surfaces/Interfaces}

Non-idealities at surfaces and/or interfaces may lead to charging of the surfaces, which can cause partial or full depletion of the charge carriers in the selenium Hall bar, especially given the high surface-to-volume ratio of the relatively thin film. The surface will have donor- and acceptor-like defects with a distribution of trap levels within the bandgap of selenium, and the net charge in these surface defects depends on the position of the Fermi-level at the interface. When the Fermi-level, $E_F$, coincides with the energy level for charge neutrality, $E_0$, the surface is charge-free. However, if $E_F$ is above (or below) $E_0$, it results in net negative (or positive) surface charge, with the amount of charge depending on the density of surface states. With a very high density of surface states, the Fermi-level becomes pinned (\textit{Fermi-level pinning}), meaning $E_F=E_0$ regardless of how much surface charge is required for charge balance. In the more general case, the net surface charge area-density $Q_s=Q_s\left( E_F-E_0\right)$ is a function of the Fermi-level. This surface charge must be balanced by the integrated net charge per area in the bulk selenium, $Q_b$, which depends on the Fermi-level, the doping density and the carrier density distribution, and the resulting band bending. Overall charge neutrality requires $Q_s+Q_b=0$, and this charge balance relation determines the Fermi-level position and thus the area density of free charge carriers in the Hall bar.

Figure \ref{fig:Figure4}(b) illustrates the energy band diagram near the surface, demonstrating how Fermi-level pinning and surface defects deplete carriers, leading to band bending and the formation of a space charge region. Figures \ref{fig:Figure4}(c) and (d) depict the resulting spatial distribution of charge carriers, either affected solely by surface defects or by both surface and interface defects. In reality, the situation likely falls somewhere between Fermi-level pinning and the gradual filling of surface states.

\begin{figure*}[t!]
    \centering
    \includegraphics[width=\textwidth,trim={0 0 0 0},clip]{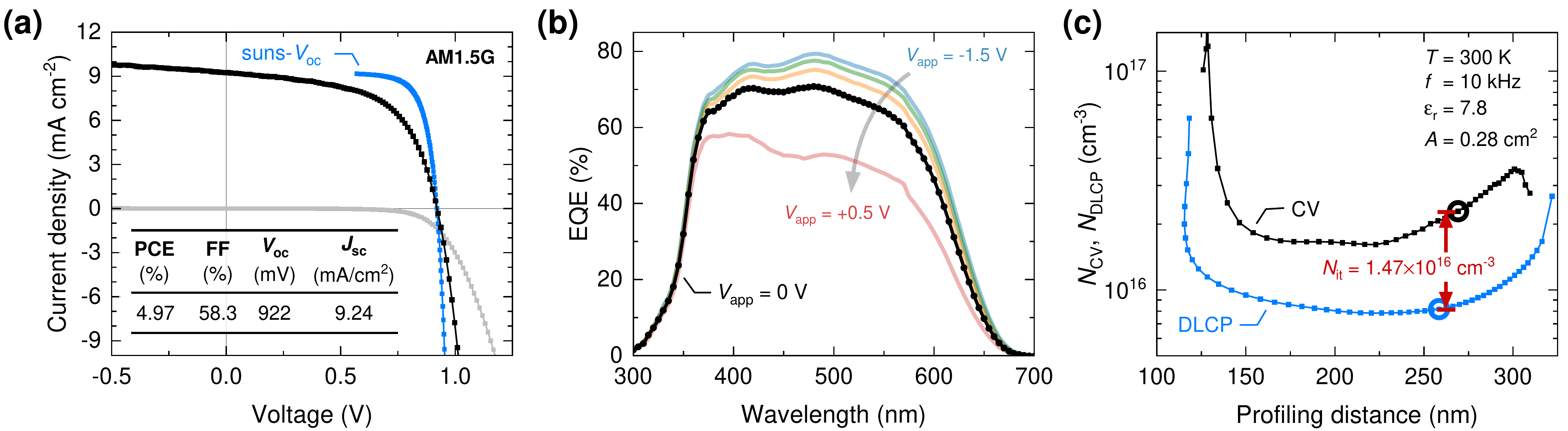}
    \caption{Photovoltaic device performance of the selenium thin-film solar cell. (a) Current density-voltage (\textit{J}--\textit{V}) curves measured under 1-sun and dark conditions, accompanied by the \textit{J}--\textit{V} curve reconstructed from suns-$V_\text{oc}$ measurements. (b) External quantum efficiency (EQE) spectrum measured under short-circuit conditions, along with EQE-spectra measured under various applied DC voltage biases. (c) Apparent carrier densities from CV and DLCP measurements at room temperature in the dark. The difference at zero bias $N_\text{it}$ presumably arises from the presence of traps at the poly-Se/ZnMgO interface.}
    \label{fig:Figure5}
\end{figure*}

\subsection{Extended Model Interpretation}

The insights gained from the temperature-dependent Hall measurements suggest that the carrier density in the selenium thin film is influenced by both incomplete ionization of acceptors and partial or full depletion due to surface and/or interface defects. Consequently, the results from carrier-resolved photo-Hall measurements may need to be reinterpreted in light of these effects.

Under low injection conditions, the spatial distribution of at least one of the charge carriers is likely to be strongly localized at the surface/interface, as illustrated in Figures \ref{fig:Figure4}(c) and (d). This localization implies that the effective mobility of the charge carriers is predominantly determined by their mobility at the non-passivated surface or interface. Moreover, the minimal overlap between the electron and hole densities under these conditions may contribute to the long recombination lifetimes.

As the injection level increases, the spatial distribution of both charge carriers broadens, allowing them to extend further into the bulk of the film. Under these high injection conditions, the effective Hall mobilities become more representative of the bulk properties, leading to an observed increase in mobility, particularly for the minority carriers. The recombination lifetime is expected to decrease with increasing carrier densities, but the greater overlap between electron and hole distributions may also contribute to the observed decrease.

With our current setup, we were only able to reach an irradiance equivalent to 0.32 suns. The observed trends show that as the injection level increases, the mobilities increases significantly. This suggests that the carrier mobilities may be higher in the bulk, with the observed mobilities of $\mu_\text{e,h}\approx 15$ cm$^\text{2}$ V$^\text{-1}$ s$^\text{-1}$ likely representing a lower bound to the mobilities under standard 1 Sun illumination, which is more relevant for photovoltaic operation. Additionally, it is important to consider that Hall measurements typically underestimate the mobilities relevant for photovoltaic devices in polycrystalline thin-films, as the lateral conductivity may be negatively affected by the presence of thousands of grain boundaries. This further supports the idea that the bulk intragrain mobilities could be higher than the observed $\mu_\text{e,h}$ values.

The carrier lifetime and diffusion lengths in selenium are observed to decrease drastically as the injection level increases, towards values more consistent with our previously reported results from time-resolved THz spectroscopy. In those measurements, two characteristic decay times of the sheet photoconductance were observed: $\tau_1=7$ ps and $\tau_2=2.2$ ns. At a pump-probe delay of 0.5 ns, the mobility sum obtained from the THz conductivity spectrum was $\Sigma \mu=5$ cm$^\text{2}$ V$^\text{-1}$ s$^\text{-1}$, which is consistent with the sum of electron and hole mobilities under low injection conditions. This comparison of the carrier properties obtained from the two different measurement techniques applied to similarly fabricated selenium thin-films on quartz substrates supports the hypothesis that at low injection levels, the measured mobilities are primarily influenced by strongly localized, long-lived carriers at the surfaces or interfaces. In contrast, at higher injection levels, the carrier-resolved photo-Hall measurements become more representative of the charge carrier properties in the bulk, with decent mobilities but very short recombination lifetimes.



\section{Photovoltaic Devices}

To contextualize the results of the temperate-dependent and carrier-resolved photo-Hall measurements, we fabricated a selenium thin-film solar cell in parallel with the selenium Hall bar. The device architecture is FTO/ZnMgO/Te/Se/MoO$_\text{x}$/Au, and the process flow is described in detail in our previous work \cite{youngman2021a, nielsen2022a, nielsen2024b}. The device was analyzed using standard electrical characterization methods, with the goal of modeling and simulating its performance based on the material properties investigated in this study.

The current density-voltage (\textit{J}--\textit{V}) characteristics of the selenium thin-film solar cell were measured under dark and 1-sun illuminated conditions, as shown in Figure \ref{fig:Figure5}(a). The cell demonstrates an efficiency of approximately 5\% and an open-circuit voltage exceeding 0.9 V, which is considered state-of-the-art for selenium-based photovoltaic devices \cite{todorov2017a, lu2024a, hadar2019a, wang2024a, liu2023a, yan2022a}. The crossover between the dark and illuminated \textit{J}--\textit{V} curves may result from changes in carrier recombination dynamics, photogenerated carrier accumulation, non-ohmic contact effects, or variations in surface states and trap behavior under illumination. However, given that the carrier mobilities are light dependent, this crossover could also simply be attributed to a significant reduction in parasitic series resistance in the photoabsorber under 1-sun conditions. The \textit{J}--\textit{V} curves are compared with a pseudo \textit{J}--\textit{V} curve reconstructed from suns-$V_\text{oc}$ measurements. Since no charge is transported during these measurements, the pseudo \textit{J}--\textit{V} curve reflects the expected \textit{J}--\textit{V} curve with zero series resistance. The difference between the pseudo fill factor ($\text{pFF}=78.0\%$) and the actual fill factor ($\text{FF}=58.3\%$) may therefore be attributed to bulk series resistance losses.

\begin{figure*}[t!]
    \centering
    \includegraphics[width=\textwidth,trim={0 0 0 0},clip]{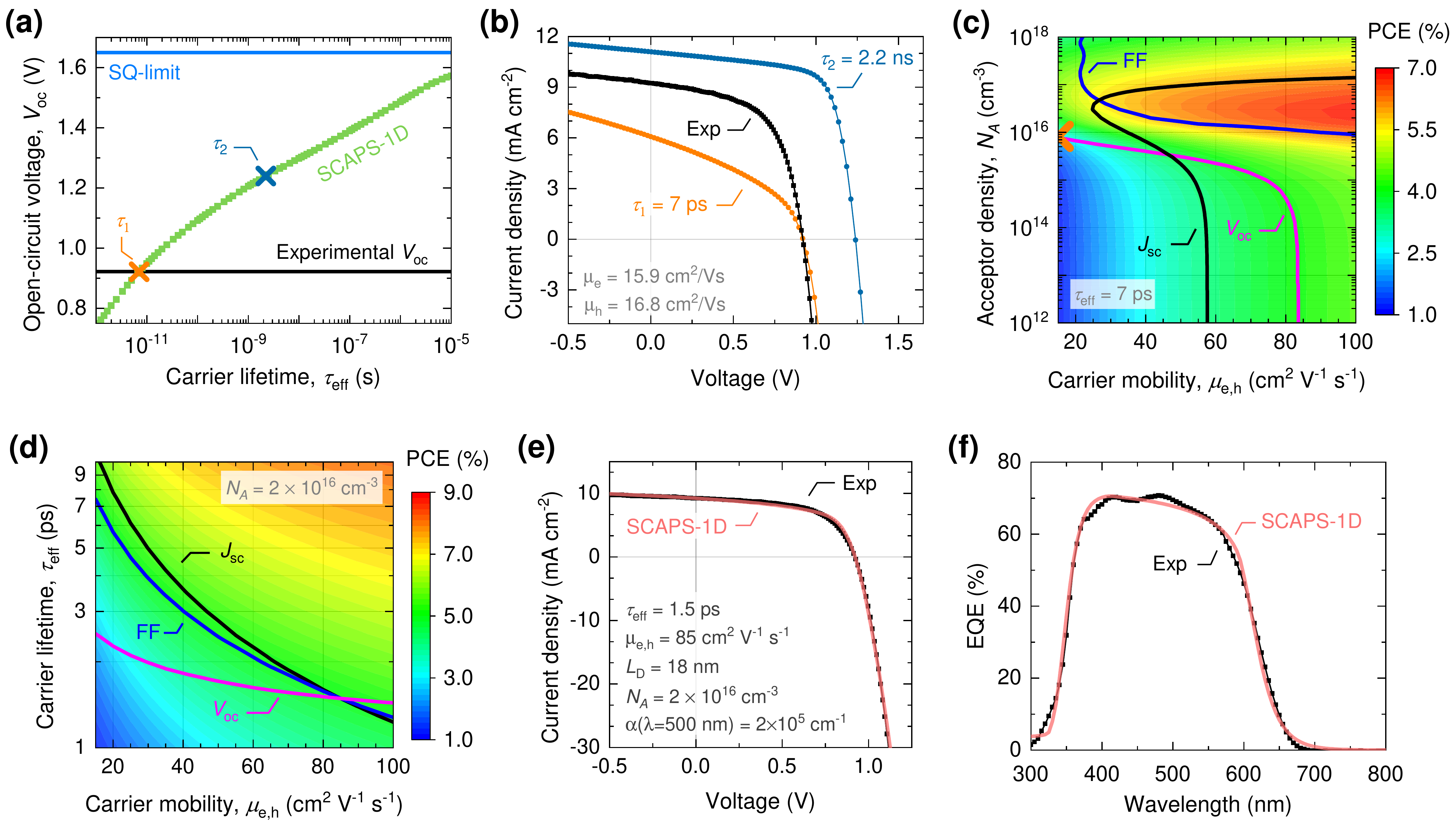}
    \caption{Photovoltaic device simulations using SCAPS-1D. (a) Open-circuit voltage ($V_\text{oc}$) as a function of the effective carrier lifetime ($\tau_\text{eff}$). The Shockley-Queisser (SQ) limit and experimental $V_\text{oc}$ are shown, with $\tau_1$ and $\tau_2$ representing the two decay times observed in THz spectroscopy transients \cite{nielsen2022a, nielsen2024b}. (b) Current density-voltage (\textit{J}--\textit{V}) curves simulated using $\tau_1$ and $\tau_2$, compared to the experimental \textit{J}--\textit{V} curve. (c) Contour plot of power conversion efficiency (PCE) as a function of acceptor density and carrier mobilities ($\mu_\text{e,h}$) with $\Delta \mu = 0$ and $\tau_\text{eff}=7$ ps. Solid lines represent matches between experimental and simulated photovoltaic metrics. (d) Contour plot of PCE as a function of $\tau_\text{eff}$ and $\mu_\text{e,h}$, with $\Delta \mu = 0$ and $N_A=2\times 10^{16}$ cm$^{-3}$. (e) The final simulated \textit{J}--\textit{V} curve, and (f) EQE spectrum compared to experimental results.}
    \label{fig:Figure6}
\end{figure*}

The external quantum efficiency (EQE) spectra shown in Figure 5(b) were measured with applied voltage biases ranging from $V_\mathrm{app} = -1.5$ V to $V_\mathrm{app} = +0.5$ V. These varying biases effectively alter the width of the depletion region in the selenium p-type absorber, as described by \cite{sze2007a, crovetto2017a}: 

\begin{equation*}
    W_p = \sqrt{\frac{2 \epsilon_0 \epsilon_{r,n} \epsilon_{r,p} \left( V_\mathrm{bi}-V_\mathrm{app}\right)N_D}{q_\mathrm{e}N_A \left( \epsilon_{r,p} N_A+ \epsilon_{r,n} N_D \right)}}
\end{equation*}

\noindent where $\epsilon_0$ is the vacuum permittivity, $\epsilon_{r,[n/p]}$ is the relative permittivity of the [n/p]-type semiconductor, $V_\mathrm{bi}$ is the built-in voltage, $q_\mathrm{e}$ is the elementary charge, $N_A$ is the acceptor density in the p-type semiconductor, and $N_D$ is the donor density in the n-type semiconductor.

The carrier collection efficiency increases under reverse bias and decreases under forward bias, indicating that the selenium photoabsorber is not fully depleted. Given the doping density of our ZnMgO electron-selective contact layer, $N_D > 10^{18}$ cm$^{-3}$ \cite{jamarkattel2022a}, a depletion region thickness less than the 300 nm thick absorber layer suggests that the doping density in selenium is on the order of $N_A \sim 10^{16}$ cm$^{-3}$. If the acceptor density in selenium were equal to the background carrier density measured in our Hall experiments ($p_0 = N_A \sim 10^{12}$ cm$^{-3}$), the depletion region width would be on the order of tens of microns, making the carrier collection efficiency in the 300 nm photoabsorber relatively insensitive to changes in the applied bias.

Finally, we carried out capacitance-voltage (CV) and drive-level capacitance profiling (DLCP) to measure the charge density at the depletion edge. The apparent doping densities are shown in Figure \ref{fig:Figure5}(c), where the plateau region corresponds to the bulk doping density of the selenium absorber. Ideally, this experiment should be performed with various absorber thicknesses to assess contributions from geometrical capacitance effects \cite{ravishankar2022a, ravishankar2021a}. Nevertheless, the results are consistent with other reports in literature and align with our hypothesis of a doping density of $N_A \sim 10^{16}$ cm$^{-3}$. As DLCP accounts for second-order contributions to the differential capacitance, the difference between DLCP and CV at $V_\mathrm{nominal}=0$ V ($N_\text{it}$) is commonly attributed to the presence of interface defects that do not respond quickly enough to the oscillating voltage signal. By multiplying $N_\text{it}$ by the depletion region width, we calculate an interface trap density of $N_\text{t} = 3.8 \times 10^{11}$ cm$^{-2}$. 
 
Both the EQE spectra and capacitance-based techniques support a doping density on the order of $N_A\sim 10^{16}$ cm$^{-3}$, consistent with our hypothesis that $p_0\neq N_A$ due to carrier freeze-out and surface/interface depletion in the Hall measurements. 

\section{Device Simulations}

The goal of this section is to evaluate whether the carrier properties investigated in this work can reproduce the experimental \textit{J}--\textit{V} curve and EQE spectrum with reasonable accuracy. For this purpose, we used the SCAPS-1D solar cell capacitance simulation tool developed by Marc Burgelman \cite{burgelman2000a}. As a starting point for the device simulations, we set the carrier mobilities to the lower bounds determined from carrier-resolved photo-Hall measurements, set the acceptor density to $N_A=10^{16}$ cm$^{-3}$, used the experimentally determined absorption coefficient from Figure \ref{fig:Figure1}(c), and incorporated the relative permittivity ($\epsilon_\text{r}$) and effective density of states ($N_{V/C}$) as determined in our earlier work \cite{nielsen2022a}. The energy band alignment, crucial for overall device performance, was thoroughly investigated in our recent study \cite{nielsen2024a}, while additional details and assumptions regarding the transport layers and ohmic contacts are provided in the Supplementary Information P.2.

Given that total current density is zero at open-circuit conditions, this operating point is less sensitive to the charge carrier mobilities, making it an ideal starting point for our analysis. First, we adjusted the effective carrier lifetime by introducing a single defect level in the selenium absorber and tuning the electron and hole capture coefficients. The simulated open-circuit voltage as a function of $\tau_\text{eff}$ is shown in Figure \ref{fig:Figure6}(a), along with the Shockley-Queisser limit of $V_\text{oc,SQ} = 1.65$ V and the experimental value of $V_\text{oc,exp} = 0.92$ V. The two decay times, $\tau_1$ and $\tau_2$, obtained from time-resolved THz spectroscopy transients, are also highlighted \cite{nielsen2022a, nielsen2024b}. Notably, the shorter decay time ($\tau_1$) provides a surprisingly good fit to the experimental \textit{V}$_{oc}$, suggesting that the effective carrier lifetime in our solar cell is likely in the picosecond regime. The longer decay time ($\tau_2$) might represent a localized carrier component with a longer lifespan, possibly indicating the time carriers remain in trap states before recombining. Figure \ref{fig:Figure6}(b) compares the simulated \textit{J}--\textit{V} curves using these two characteristic decay times to the experimental curve, where the shorter decay time yields a better fit. This becomes more apparent when considering that we used the conservative lower bounds for the carrier mobilities, which, if increased, would improve both the fill factor and short-circuit current density.

To refine our analysis, we fixed the effective carrier lifetime at $\tau_\text{eff} = 7$ ps, based on the better fit provided by the shorter THz transient decay time. We then varied the acceptor density and carrier mobilities. As the carrier-resolved photo-Hall measurements indicate that $\Delta \mu_\text{Hall}$ converges toward zero with increasing illumination, we assumed equal electron and hole mobilities ($\mu_\text{e} = \mu_\text{h}$) and only investigated values greater than 15 cm$^\text{2}$ V$^\text{-1}$ s$^\text{-1}$. The results are shown in Figure \ref{fig:Figure6}(c) as a contour plot of the power conversion efficiency (PCE). Here, the solid lines represent a match between the experimental and simulated values of key photovoltaic metrics, including the fill factor ($\text{FF}$), short-circuit current density (\textit{J}$_{sc}$), and open-circuit voltage (\textit{V}$_{oc}$). Notably, the experimental $\text{FF} = 58.3\%$ cannot be reproduced with an acceptor density of $N_A < 10^{16}$ cm$^{-3}$, supporting our hypothesis that $p_0 \neq N_A$.

Since the three solid lines in Figure \ref{fig:Figure6}(c) do not intersect, no unique solution fits the \textit{J}--\textit{V} curve perfectly within the current model constraints. Therefore, we fixed the acceptor density at $N_A = 2 \times 10^{16}$ cm$^{-3}$, as determined from the CV measurements in Figure \ref{fig:Figure5}(c). We then varied the effective carrier lifetime and mobilities, with the results presented in Figure \ref{fig:Figure6}(d) as a contour plot. This plot reveals a broad range of mobility-lifetime combinations that result in a PCE of approximately 5\%, with reasonable fits for both the fill factor and short-circuit current density. Finally, all three solid lines intersect at a single point, corresponding to an effective carrier lifetime of $\tau_\text{eff} = 1.5$ ps and carrier mobilities of $\mu_\text{e,h} = 85$ cm$^2$ V$^{-1}$ s$^{-1}$. The simulated and experimental \textit{J}--\textit{V} curves and EQE spectra are compared in Figures \ref{fig:Figure6}(e) and (f).\\

From our device simulation study, we conclude that the shorter of the two characteristic decay times from the time-resolved THz spectroscopy transients should be considered the effective carrier lifetime. This finding implies that the carrier dynamics limiting the experimental photovoltaic performance are within the picosecond regime. Additionally, the mobilities of these short-lived carriers are significantly higher than first anticipated. While we had considered $\mu_\text{e,h} \approx 15$ cm$^2$ V$^{-1}$ s$^{-1}$ as a conservative lower bound, higher effective mobilities are reasonable when accounting for the influence of thousands of grain boundaries on Hall effect measurements, the observed increase in mobility with higher illumination levels, and the fact that we only measured up to 0.3 suns. 

There is a discrepancy between the mobility sum derived from THz measurements and the photo-Hall mobilities, but given that the THz sheet photoconductivity spectrum was obtained with a pump-probe delay of 0.5 ns, this sum reflects the properties of the already localized charge carriers. Therefore, future time-resolved THz spectroscopy studies on selenium should focus on the carrier dynamics within the first few picoseconds post-excitation. Additionally, using wavelength-dependent optical pumps could help resolve carrier mobilities spatially, thereby enhancing our understanding of surface defects. It is important to note that optical pump THz spectroscopy transients and photo-Hall measurements capture carrier dynamics under different conditions. While photo-Hall measurements reflect steady-state carrier mobilities under illuminated conditions, THz transients measure dynamic responses, emphasizing the need for complementary approaches to fully understand carrier behavior in photovoltaic devices.

Finally, the diffusion length $L_\text{D} = 18$ nm, as determined from our device simulations, is shorter than previously estimated. However, given that the depletion region width under short-circuit conditions is approximately 250–275 nm based on our capacitance-based measurements, our conclusion remains that an absorber thickness of around $\sim 300$ nm is optimal given the optoelectronic quality of our selenium thin-films.






\section{Conclusion}

In summary, we carried out carrier-resolved photo-Hall measurements on a high-performance selenium thin-film, providing detailed insights into the behavior of both majority and minority carriers under varying light conditions. Given the unexpectedly low carrier densities in the dark relative to literature values for the doping/acceptor density, we also conducted temperature-dependent Hall measurements. These measurements revealed significant evidence of carrier freeze-out and depletion through surface and/or interface defects, suggesting that at low injection levels, the carrier properties may be influenced by strong surface localization effects. Consequently, under low injection conditions, our results may reflect the effective mobilities at the surface or interface rather than in the bulk, whereas the bulk mobilities become more apparent as the carrier distribution broadens at higher injection levels. Additionally, the derived recombination lifetime appears high due to the minimal overlap of the carrier distribution profiles at low injection, and reduces dramatically with increasing illumination as the overlap becomes more pronounced.

To critically assess our results, we fabricated a selenium thin-film solar cell in parallel with the Hall bar, and analyzed the device using standard electrical characterization techniques. Device simulations were then used to systematically evaluate each material and carrier property, allowing us to accurately reproduce experimental current-voltage and external quantum efficiency measurements. Our simulations suggest that the effective carrier lifetime relevant for photovoltaic operation is in the picosecond range, rather than the nanosecond range previously reported, and the bulk carrier mobilities are significantly higher than earlier estimates derived from transient THz spectroscopy. This discrepancy we attribute to the 0.5 nanosecond pump-probe delay in the THz experiment, which consequently measures the mobility sum of the already localized, short-lived carriers. Insights from our photo-Hall experiments, supported by detailed device simulations, indicate that while carrier lifetimes are in the picosecond regime, the intrinsic material properties of selenium are highly promising. The observed open-circuit voltage deficit is primarily due to shorter carrier lifetimes than previously reported, with the longer-lived decay components of THz transients likely resulting from trapping/localization of charge carriers. These findings imply a high intrinsic photovoltaic performance potential for selenium photoabsorbers and underscore the need for defect engineering to reduce non-radiative recombination losses and achieve higher efficiency selenium solar cells.

\section*{Methods}

\paragraph*{\normalfont \textbf{Materials:}} Amorphous selenium (99.999+\%, metals basis) and tellurium (99.9999\%, metals basis) shots were purchased from Alfa Aesar. Fused quartz plates (20$\,$x$\,$20$\,$x$\,$0.5$\,\,$mm) were purchased from Machined Quartz. The Au (99.99\%) sputtering target was purchased from AJA International.\\

\paragraph*{\normalfont \textbf{Fabrication of selenium Hall samples:}} Quartz substrates are sequentially cleaned for 15 minutes each in Milli-Q water (18.2 M$\Omega\,$cm$\,$@$\,$25$^\circ$C), acetone and isopropanol using ultrasound. The cleaned substrates are then dried using a nitrogen gun and transferred to a custom-built thermal evaporator at a base pressure of $P_\text{base}\sim\text{10}^\text{-8}$ mbar. A dedicated shadow mask is placed in intimate contact with the substrates to form a Hall bar with an active length of $L=\text{4}$ mm and width of $W=\text{2}$ mm. First, $\sim\text{1}$ nm of tellurium is evaporated at a rate of 0.3 Å/s, followed by the evaporation of $\approx\text{300}$ nm selenium at a rate of 3.0 Å/s. The samples are then thermally annealed on a hotplate in air at 190$^\circ$C for 4 minutes to crystallize the selenium Hall bar. Finally, a 500 nm thick Au contact grid is sputter deposited onto the Hall bar through a dedicated shadow mask using 50 W DC power at 404 V, an Ar flow of 30 sccm and 5 mTorr pressure, and substrate rotation to improve uniformity. The fabrication process flow is schematically illustrated in Fig \ref{fig:Figure1}(a).\\

\paragraph*{\normalfont \textbf{Carrier-resolved photo-Hall measurements:}} A high-sensitivity AC Hall measurement system was used, based on a set of rotating parallel dipole line (PDL) magnets \cite{gunawan2015a}. The PDL master magnet was rotated by a stepper motor system at a speed of 1.25 r.p.m. to generate an oscillating magnetic field with a peak amplitude of 0.5 T \cite{gunawan2015patent}. Photoexcitation was achieved using solid-state laser ($\lambda = 520$ nm), with the emitted laser beam guided through motorized neutral density filters. The incident photon flux was determined using a beam splitter, which directed part of the beam to a calibrated silicon photodiode at each light intensity. A Keithley 2450 source meter was used for current sourcing, and a Keithley 2001 digital multimeter for voltage measurements. The photodiode current was measured using a Keithley 617 electrometer. Temperature-dependent Hall measurements were carried out under dark conditions using a Displex cryostat system. The Hall signal analysis was performed by Fourier transforming the magnetoresistance measurements and employing lock-in detection to extract the component in phase with the oscillating magnetic field, eliminating out-of-phase contributions, such as those caused by Faraday induction of an electromotive force \cite{gunawan2017patent}. Additional details on the photo-Hall setup and analysis are elaborated in Ref \cite{gunawan2019a}.\\

\paragraph*{\normalfont \textbf{Device characterization:}} Current-voltage (\textit{I}--\textit{V}) measurements of photovoltaic devices were performed using a Keithley 2561A source meter with 4-terminal sensing under 1 sun illumination (Newport 94082A solar simulator, class ABA, equipped with a 1600 W Xe arc lamp and appropriate AM1.5G filters). The light intensity was calibrated in the plane of the device under test using a reference solar cell from Orion. As no mask aperture was used during the acquisition, the active area was determined by calculating the AM1.5G equivalent current density using the integral of the external quantum efficiency (EQE) spectrum of the device under test. EQE spectra were measured using the QEXL from PV Measurements, calibrated with a silicon reference photodiode. For capacitance-voltage (CV) measurements, the impedance spectra were recorded using a BK895 LCR meter operating in resistance-reactance mode. The measurements were carried out across a frequency range from 0.1 kHz to 1 MHz with an AC level of 30 mV, and DC biases ranged from -1.0 V to +1.5 V. For drive-level capacitance profiling (DLCP), the nominal DC voltage was also varied from -1.0 V to +1.5 V, with AC levels ($V_\text{peak}$) between 2 and 70 mV. Both the CV and DLCP measurements were conducted at a frequency of 10 kHz at room temperature in the dark. Suns-$V_\text{oc}$ measurements are conducted using the accessory stage to a WCT-120 from Sinton Instruments. The setup uses an illumination sensor calibrated from 0.006 to 6 suns, and the sample chuck is temperature controlled at 25$^\circ$C.\\

\paragraph*{\normalfont \textbf{Additional characterization:}} Scanning electron microscopy (SEM) images are obtained using the Supra 40 VP SEM from Zeiss. The absorption coefficient is derived from Lambert Beer's law using the thickness of the film, and reflection-corrected UV-vis transmission data, $\textit{T}_{\text{corr}}=\textit{T}/\left(1-R\right)$, measured at room temperature over a $\sim$2$\pi$ sr solid angle using a double-beam Cary 7000 spectrophotometer equipped with an integrating sphere.\\

\section*{Conflicts of interest}
There are no conflicts of interest to declare.\\

\section*{Acknowledgements}
The work presented here is supported by the Carlsberg Foundation, grant CF24-0200, and the Independent Research Fund Denmark (DFF) grant 0217-00333B. R.S.N. would like to thank Dan Shacham for his assistance with the dedicated sample holders and shadow masks for the fabrication of the Hall bar and the Au contact grid. \\

\section*{Data availability}
The data that support the findings of this study are available from the corresponding author upon request.\\


\nocite{*}

\bibliography{references}

\end{document}



\title{\Large{SUPPORTING INFORMATION} \\ \vspace{1cm} \large Variable Temperature and Carrier-Resolved Photo-Hall Measurements of High-Performance Selenium Thin-Film Solar Cells}

\author{Rasmus S. Nielsen}
\email[]{Electronic mail: rasmus.nielsen@empa.ch}
\affiliation{Transport at Nanoscale Interfaces Laboratory, Swiss Federal Laboratories for Material Science and Technology (EMPA), Ueberlandstrasse 129, 8600 Duebendorf, Switzerland}

\author{Oki Gunawan}
\affiliation{IBM Thomas J. Watson Research Center, 1101 Kitchawan Road, Yorktown Heights, NY 10598, USA}

\author{Teodor Todorov}
\affiliation{IBM Thomas J. Watson Research Center, 1101 Kitchawan Road, Yorktown Heights, NY 10598, USA}

\author{Clara B. Møller}
\affiliation{SurfCat, DTU Physics, Technical University of Denmark, 2800 Kongens Lyngby, Denmark}

\author{Ole Hansen}
\affiliation{National Center for Nano Fabrication and Characterization (DTU Nanolab), Technical University of Denmark, 2800 Kongens Lyngby, Denmark}

\author{Peter C. K. Vesborg}
\affiliation{SurfCat, DTU Physics, Technical University of Denmark, 2800 Kongens Lyngby, Denmark}

\maketitle

\vfill

\tableofcontents

\clearpage

\section{Derivation of acceptor ionization efficiency}

In a charge neutral p-type semiconductor with an acceptor density $N_A$ at the acceptor level $E_A$, and thus $\Delta E_A=E_A-E_\mathrm{VBM}$ where $E_\mathrm{VBM}$ is the top of the valence band, and acceptor degeneracy $g_A$ (mostly $g_A=4$), the density of ionized acceptors is

\begin{align*}
    N_A^-&=\frac{N_A}{1+g_A\exp{\left(-\frac{E_F-E_A}{k_\mathrm{B}T}\right)}}=\frac{N_A}{1+g_A\frac{p}{p_1}},\text{ with }\\
    p&=N_V\exp{\left(-\frac{E_F-E_\mathrm{VBM}}{k_\mathrm{B}T}\right)},\text{ and }p_1=N_V\exp{\left(-\frac{E_A-E_\mathrm{VBM}}{k_\mathrm{B}T}\right)}
\end{align*}

\noindent Charge neutrality requires $p-n-N_A^-+N_D^+=0$ which with zero compensation ($N_D^+=0$) and sufficiently large acceptor density may be approximated (i.e. neglect $n$) to

\begin{align*}
    p &\simeq N_A^-=\frac{N_A}{1+g_A\frac{p}{p_1}} \\
    p+g_A\frac{p^2}{p_1}-N_A &\simeq 0
\end{align*}

\noindent with the solution 

\begin{align*}
    \frac{p}{N_A} = \frac{N_A^-}{N_A} = \frac{\sqrt{1+4\left[\frac{g_AN_A}{p_1}\right]}-1}{2\left[\frac{g_AN_A}{p_1}\right] } = \frac{\sqrt{1+4g_A \left[\frac{N_A}{N_V}\right] \exp{\left(\frac{\Delta E_A}{k_\mathrm{B}T}\right)}}-1}{2g_A \left[\frac{N_A}{N_V}\right] \exp{\left(\frac{\Delta E_A}{k_\mathrm{B}T}\right)}}
\end{align*}

\noindent with the two limiting cases, full ionization and freeze-out, respectively

\begin{align*}
    \frac{p}{N_A} &\rightarrow 1,\text{ for }4g_A\left[\frac{N_A}{N_V}\right]\exp{\left(\frac{\Delta E_A}{k_\mathrm{B}T}\right)} \ll 1\\
    \frac{p}{N_A} &\rightarrow \sqrt{\frac{N_V}{g_AN_A}}\exp{\left(-\frac{\Delta E_A}{2k_\mathrm{B}T}\right)},\text{ for }4g_A\left[\frac{N_A}{N_V}\right]\exp{\left(\frac{\Delta E_A}{k_\mathrm{B}T}\right)} \gg 1
\end{align*}

\clearpage

\section{SCAPS-1D device simulations}

$\textbf{Contact properties:}\quad$ The surface recombination velocities on both the front and back contacts were assumed to be $S_\mathrm{e,h}=10^7$ cm/s, representing surfaces with infinitely fast recombination. The work function of the FTO contact at the front and the Au contact at the back of our device were determined to be $\phi=4.32$ eV and $\phi=5.20$ eV, respectively, based on our previous work using voltage-biased UV photoemission spectroscopy (UPS) \cite{nielsen2024a}. Given the thickness of $d_\text{Au}>50$ nm, we assume the Au contact is 100\% reflective, whereas the FTO contact was approximately 80\% transparent, according to specifications from Sigma-Aldrich.

$\textbf{FTO:}\quad$ The FTO contact was explicitly defined as a layer in the device structure to account for parasitic absorption, band tailing, finite carrier mobilities, and particularly, contact tunneling of electrons through a transport barrier to the electron-selective contact, ZnMgO. Therefore, intraband tunneling at the FTO/ZnMgO interface was enabled. The sheet resistance of the commercial FTO-coated glass was specified as 7 $\Omega$/sq, and based on the thickness estimated from cross-sectional SEM images ($d_\mathrm{FTO}\approx500$ nm) and an electron mobility of 22 cm$^\text{2}$ V$^\text{-1}$ s$^\text{-1}$ (considering the heat treatments involved in our device processing) \cite{luangchaisri2012a}, the doping density was calculated to be $N_D=8.1 \times 10^{20}$ cm$^{-3}$. Since the device simulations were relatively insensitive to the dielectric permittivity $\epsilon_\text{r}$, effective density of states $N_{C,V}$, and minority carrier mobilities of this layer, we used the reported values from the SCAPS-1D simulation study of CZTS-based solar cells by S. H. Zyoud et al. \cite{zyoud2021a}.

$\textbf{ZnMgO:}\quad$ The sputtering target used for synthesizing this layer had a composition of Zn$_\text{0.85}$Mg$_\text{0.15}$O. The target composition, materials supplier, sputtering conditions, and the subsequent high-vacuum annealing process were similar to those reported by Jamarkattel et al. \cite{jamarkattel2022a}. In their study, the ZnMgO thin film was characterized using Hall effect measurements to determine the doping density and majority carrier mobility. Given the spike-like alignment of the conduction band edges with the selenium absorber \cite{nielsen2024a}, intraband tunneling was enabled at the ZnMgO/Se interface. Additionally, a single-level defect was introduced with a trap density, determined by quantitative comparison of CV- and DLCP-measurements, $N_\mathrm{t}=3.8 \times 10^{11}$ cm$^{-2}$. The thickness, optical bandgap ($E_\mathrm{g}$), electron affinity ($\chi$), and crystallization of ZnMgO in the wurtzite structure were established in our previous work \cite{nielsen2024a}. Based on the elemental composition and crystal structure, effective masses ($m_\mathrm{e}=0.23 \, m_0$ and $m_\mathrm{h}=0.255 \, m_0$) were adopted from Franz et al. \cite{franz2013a} Using the parabolic band approximation, the effective density of states was calculated as:

\begin{equation*}
    N_{C,V}=2\left( \frac{2\pi m_\mathrm{e,h}^* k_\mathrm{B}T}{h^2} \right)
\end{equation*}

\noindent where $h$ is Planck's constant, $k_\mathrm{B}$ is the Boltzmann constant and $T$ is the temperature.

$\textbf{poly-Se:}\quad$ 
The material properties of a polycrystalline selenium thin-film, fabricated similarly to the one studied in this work, were investigated in great details in our previous publications \cite{nielsen2022a, nielsen2024a}. A single-level neutral defect was defined in the bulk of the selenium thin-film, with the product of the defect density $N_\mathrm{t}$ and the capture cross-sections $\sigma_\mathrm{e,h}$ (assumed to be similar for both carriers) varied to adjust the effective carrier lifetime. The lifetime is inversely proportional to the $N_\mathrm{t}\times\sigma_\mathrm{e,h}$ product, as given by:

\begin{equation*}
    \tau_\mathrm{e,h} = \frac{1}{N_\mathrm{t}\sigma_\mathrm{e,h}\nu_\mathrm{th,(e,h)}}
\end{equation*}

\noindent where $\nu_\mathrm{th,(e,h)}$ is the thermal velocity of the charge carriers.

$\textbf{MoO$_\text{x}$:}\quad$ MoO$_\text{x}$ is challenging to define in SCAPS-1D using its real material properties, as its high work function relative to the adjacent layers causes convergence issues in the software. The high work function is beneficial as it sets up a back surface field/induces band bending, repelling electrons and attracting holes. Carrier transport through the MoO$\text{x}$ layer is described by a defect band formed by oxygen vacancies in stoichiometric MoO$_3$ \cite{nakayama2009a, battaglia2014a}. To circumvent the high work function issue, this defect band may be defined as the valence band edge with a low effective density of states and a smaller bandgap. We used values from W. Li et al. \cite{li2019a}, where the back interface of high-efficiency chalcopyrite solar cells was numerically investigated, resulting in accurate curve fits to experimental \textit{J}--\textit{V} curves for both CIGS and various perovskite-based solar cells. The thickness was determined using variable-angle spectroscopic ellipsometry.\\

Finally, the parasitic series resistance was set to $R_\mathrm{s}=3.26 \, \Omega $ cm$^2$, as determined from the real part of the impedance at 1 MHz. The shunt resistance was not explicitly defined but can be estimated to be $R_\mathrm{sh}\approx1100\,\Omega $ cm$^2$ from the slope of the experimental \textit{J}--\textit{V} curves at short-circuit conditions. All absorption coefficients were defined using experimental data from UV-vis or spectroscopic ellipsometry.

\clearpage

\begin{table}[h!]
\small
  \caption{\ Material parameters used in SCAPS-1D device simulations.}
  \vspace{0.25cm}
  \label{tbl:example}
  \begin{tabular*}{\textwidth}{@{\extracolsep{\fill}}lrrrr}
    \hline \vspace{-0.25cm}\\
    \textbf{Contact properties} & \textbf{Front} & & & \textbf{Back} \\
    $S_\mathrm{e}$ (cm/s) & $\text{10}^{\text{7}}$ & & & $\text{10}^{\text{7}}$ \\
    $S_\mathrm{h}$ (cm/s) & $\text{10}^{\text{7}}$ & & & $\text{10}^{\text{7}}$ \\
    $\phi$ (eV) & 4.32 (flat bands) & & & 5.20 \\
    $T/R$ (\%)\vspace{0.2cm} & 80/0 & & & 0/100 \\
    \hline \vspace{-0.25cm}\\
    \textbf{Layer properties}$^\star$ & \textbf{FTO} & \textbf{ZnMgO} & \textbf{poly-Se}$^*$ & \textbf{MoO$_\text{x}$} \cite{li2019a} \\
    Thickness (nm) & $\text{500}$ & $\text{65}$ & $\text{300}$ & $\text{15}$ \\
    $E_\mathrm{g}$ (eV) & 3.80 \cite{banyamin2014a} & $\text{3.55}$ \cite{nielsen2024a} & $\text{1.95}$ \cite{nielsen2022a} & 3.00 \\
    $\chi$ (eV) & 4.44 & 3.69 \cite{nielsen2024a} & 3.89 \cite{nielsen2024a} & 2.50 \\
    $\epsilon_\mathrm{r}$ & 9.0 \cite{zyoud2021a} & 8.0 \cite{ashkenov2003a} & 7.8 \cite{nielsen2022a} & 12.5 \\
    $N_C$ (cm$^{-3}$) & $\text{1.00}\times \text{10}^{\text{18}}$ \cite{zyoud2021a} & $\text{2.77}\times \text{10}^{\text{18}}$ \cite{franz2013a} & $\text{8.70}\times \text{10}^{\text{18}}$ \cite{nielsen2022a} & $\text{2.20}\times \text{10}^{\text{18}}$ \\
    $N_V$ (cm$^{-3}$) & $\text{1.00}\times \text{10}^{\text{18}}$ \cite{zyoud2021a} & $\text{3.23}\times \text{10}^{\text{18}}$ \cite{franz2013a} & $\text{1.64}\times \text{10}^{\text{20}}$ \cite{nielsen2022a} & $\text{1.80}\times \text{10}^{\text{19}}$ \\
    $\mu_\mathrm{e}$ (cm$^{2}$/Vs) & 22 \cite{luangchaisri2012a} & 1.5 \cite{jamarkattel2022a} & 85$^\ddag$ & 25 \\
    $\mu_\mathrm{h}$ (cm$^{2}$/Vs) & 10 \cite{zyoud2021a} & 1.5$^\dag$ & 85$^\ddag$ & 100 \\
    $N_{D/A}$ (cm$^{-3}$)\vspace{0.2cm} & $N_D= \text{8.1} \times \text{10}^{\text{20}}$ & $N_D= \text{3.2} \times \text{10}^{\text{18}}$ \cite{jamarkattel2022a} & $N_A= \text{2} \times \text{10}^{\text{16}}$ & $N_A= \text{6.0} \times \text{10}^{\text{18}}$ \\
    \hline \vspace{-0.25cm}\\
    \textbf{Defect states} & \textbf{FTO} & \textbf{ZnMgO} & \textbf{poly-Se}$^\mathsection$ & \textbf{MoO$_\text{x}$} \\
    Type & - & - & Neutral & - \\
    Energy distribution & - & - & Single level & - \\
    $N_\mathrm{t}$ (cm$^{-3}$) & - & - & $\text{1.5} \times \text{10}^{\text{16}}$ & - \\
    $E_\mathrm{t}$ (eV) & - & - & $\textit{E}_\text{V} + \text{0.6}$ & - \\
    $\sigma_\mathrm{e}$, $\sigma_\mathrm{h}$ (cm$^{2}$)\vspace{0.2cm} & - & - & $\text{4.5} \times \text{10}^{-\text{12}}$ & - \\
    \hline \vspace{-0.25cm}\\
    \textbf{Interface defect} & \textbf{FTO/ZnMgO} & \textbf{ZnMgO/Se} & \textbf{poly-Se/MoO$_\text{x}$} & \\
    Type & - & Neutral & - & \\
    Energy distribution & - & Single level & - & \\
    $E_\mathrm{t}$ (eV) & - & $\textit{E}_\text{V} + \text{0.6}$ & - & \\
    $N_\mathrm{t}$ (cm$^{-2}$) & - & $\text{3.8} \times \text{10}^{\text{11}}$ & - & \\
    $\sigma_\mathrm{e}$, $\sigma_\mathrm{h}$ (cm$^{2}$) & - & $\text{10}^{-\text{15}}$ & - & \\
    Intraband tunneling & Enabled & Enabled & - & \\
    \qquad \vspace{-0.25cm} \\
    \hline \vspace{-0.25cm}\\
  \end{tabular*}
  \begin{minipage}{\textwidth}
    \raggedright
    \small $^\star$ \footnotesize The thermal velocity of electrons and holes in all layers are assumed to be $\nu_\mathrm{th,(e,h)}=10^7$ cm/s.\\
    \small $^*$ \footnotesize The material properties of selenium, where referenced, are from our previously published works and therefore accurately represent the optoelectronic quality of selenium thin-films studied in this work \cite{nielsen2022a}, synthesized using the same custom tools and in-house recipes. \\ 
    \small $^\dag$ \footnotesize The minority carrier (hole) mobility is assumed to be similar to the majority carrier (electron) mobility in ZnMgO as determined by dark Hall-effect measurements \cite{jamarkattel2022a}.\\
    $^\ddag$ The carrier mobilities in selenium are assumed to be similar under illuminated conditions given the convergence of the Hall-mobility difference $\Delta \mu_\text{Hall} \rightarrow 0$ with increasing absorbed photon density $G_\gamma$.\\
    \small $^\mathsection$ \footnotesize The effective carrier lifetime in selenium is adjusted by varying the capture cross sections, assumed to be similar for both majority and minority carriers, while maintaining a fixed defect density.
  \end{minipage}
  \vspace{-0.4cm}
\end{table}

\clearpage

\subsection{Acceptor density VS carrier mobility}

\begin{figure*}[h!]
    \centering
    \includegraphics[width=\textwidth,trim={0 0 0 0},clip]{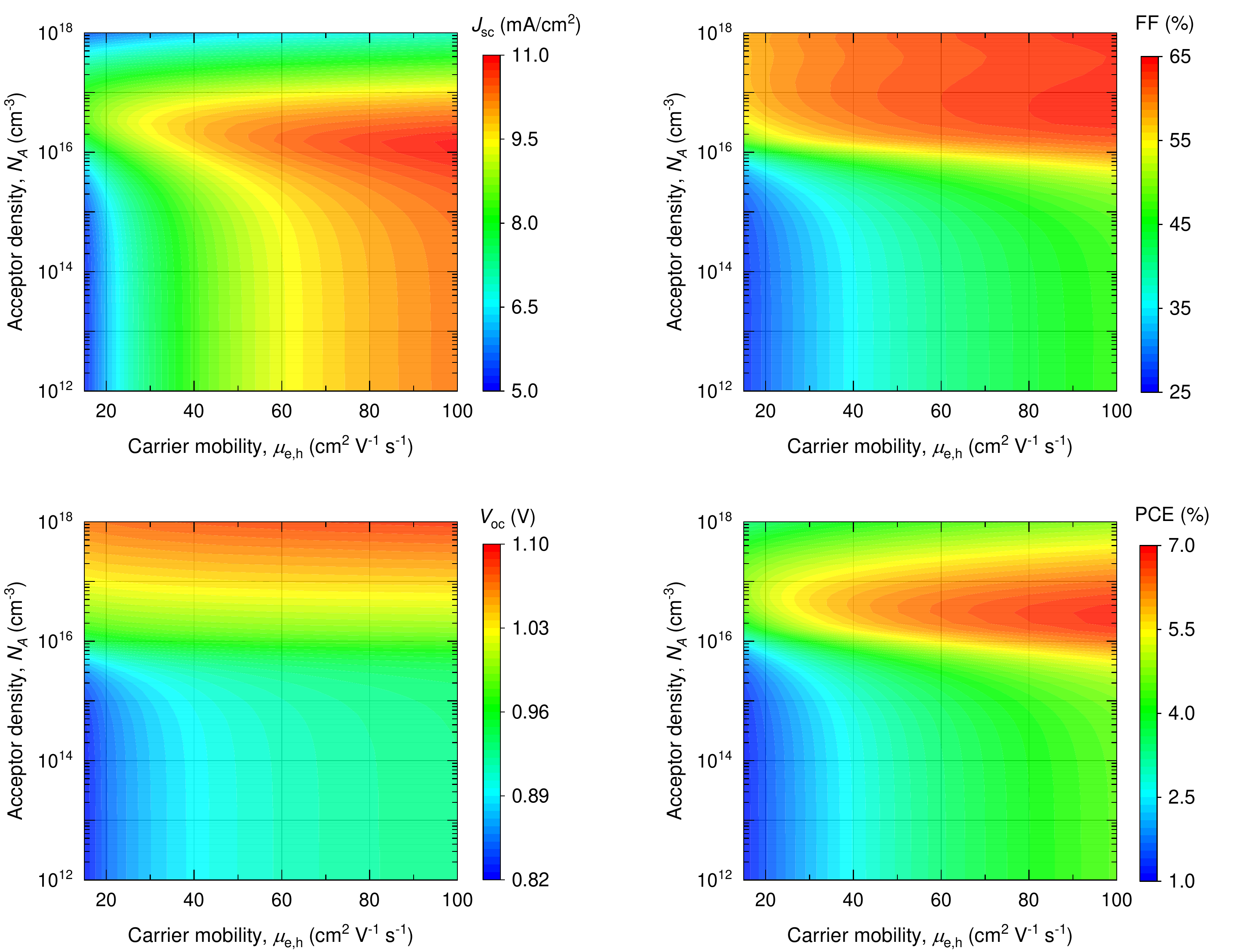}
    \caption{Contour plots of the individual photovoltaic performance metrics within the parameter space of acceptor density and carrier mobility. An effective carrier lifetime of $\tau_\text{eff}=7$ ps is assumed.}
    \label{fig:ESI_Figure7}
\end{figure*}

\clearpage

\subsection{Carrier lifetime VS carrier mobility}

\begin{figure*}[h!]
    \centering
    \includegraphics[width=\textwidth,trim={0 0 0 0},clip]{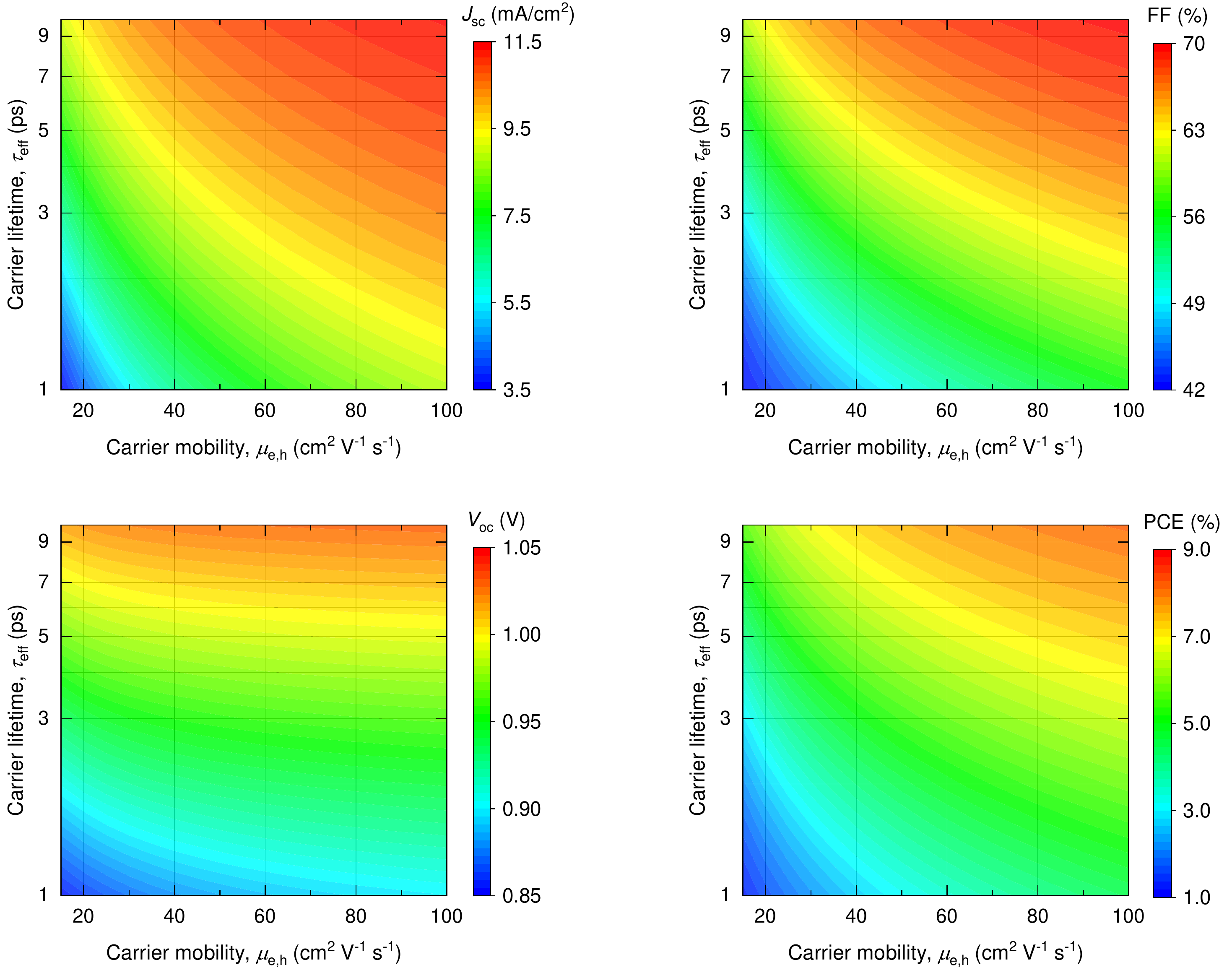}
    \caption{Contour plots of the individual photovoltaic performance metrics within the parameter space of carrier lifetime and carrier mobility. An acceptor density of $N_A=2\times10^{16}$ cm$^{-3}$ is assumed.}
    \label{fig:ESI_Figure8}
\end{figure*}

\clearpage

\subsection{Energy band diagrams and spatially resolved current densities}

\begin{figure*}[h!]
    \centering
    \includegraphics[width=\textwidth,trim={0 0 0 0},clip]{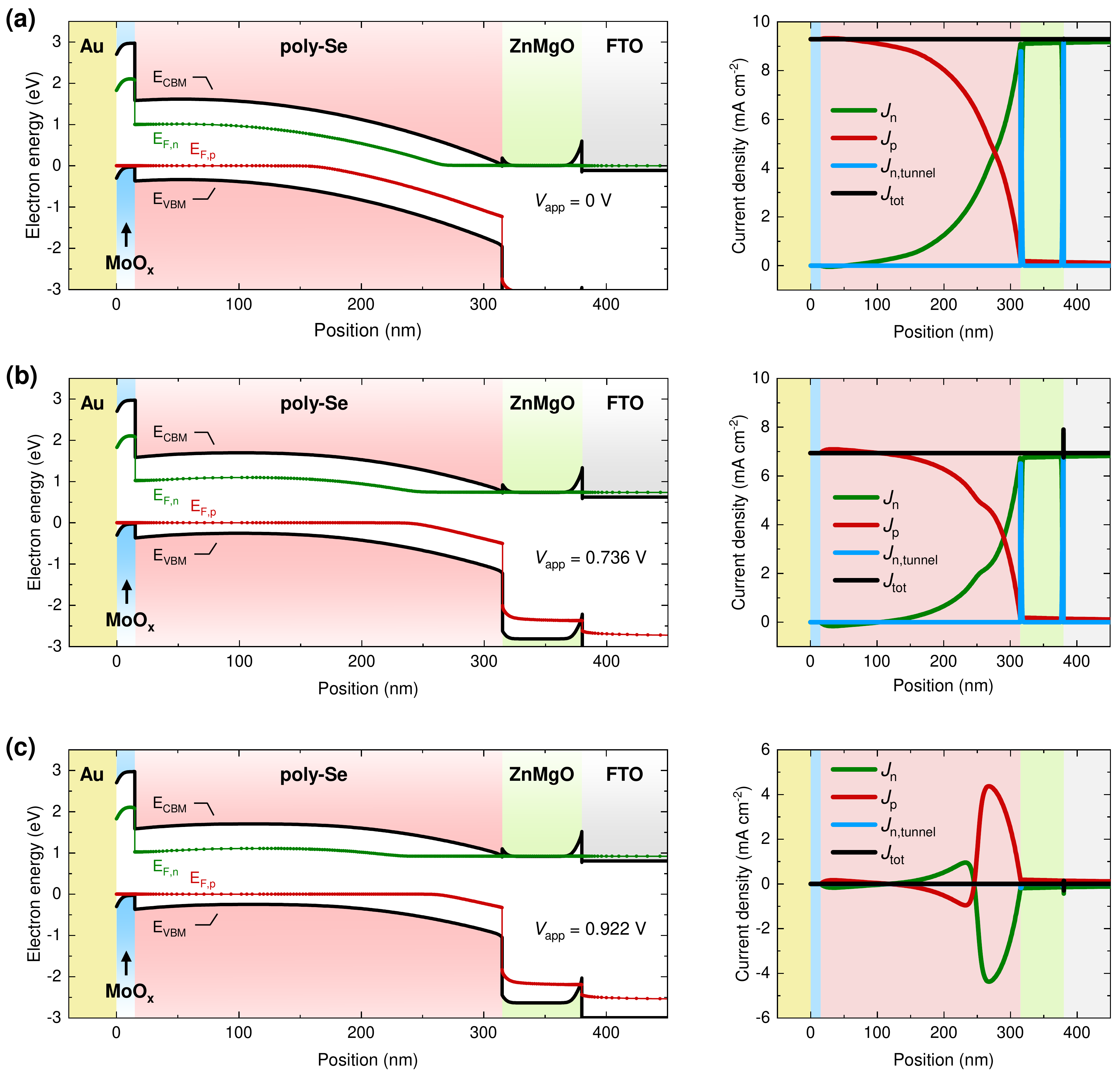}
    \caption{Simulated energy band diagrams of the selenium thin-film solar cell under (a) short-circuit, (b) maximum-power, and (c) open-circuit conditions. The corresponding current densities are shown on the right.}
    \label{fig:ESI_Figure10}
\end{figure*}

\clearpage

\section{Carrier mobility under 1-sun illumination}

\begin{figure*}[h!]
    \centering
    \includegraphics[width=0.48\textwidth,trim={0 0 0 0},clip]{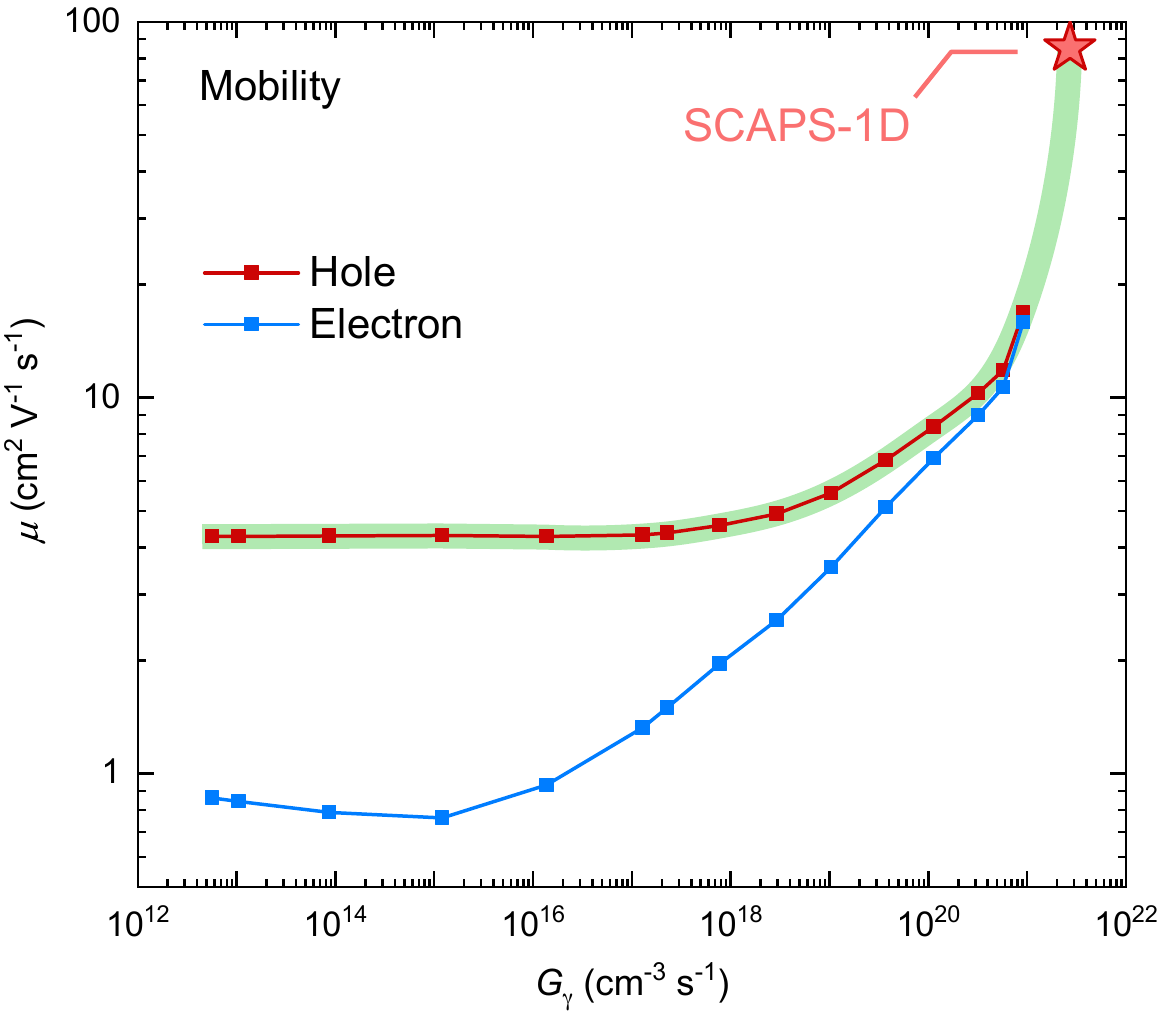}
    \caption{Carrier-resolved Hall mobilities from dark to 0.3 suns equivalent illuminated conditions, extrapolated to the predicted value under 1-sun conditions from SCAPS-1D device simulations. It is important to note that the photo-Hall measurements may be negatively influenced by thousands of grain boundaries in the polycrystalline thin-film, and thus potentially underestimating the effective carrier mobility relevant for photovoltaic operation.}
    \label{fig:ESI_Figure9}
\end{figure*}

\clearpage

\section{Capacitance-voltage (CV) measurements}

\begin{figure*}[h!]
    \centering
    \includegraphics[width=0.8\textwidth,trim={0 0 0 0},clip]{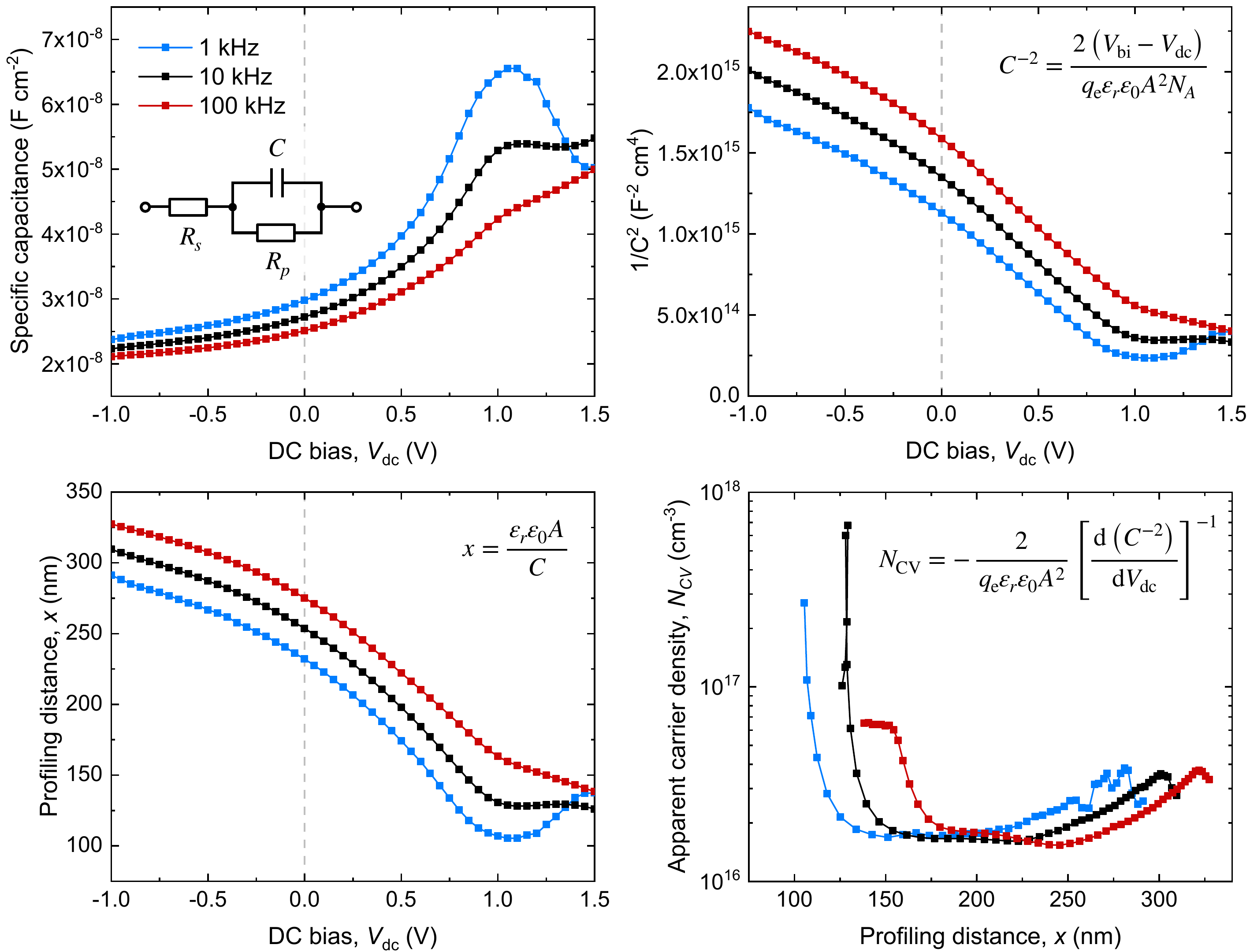}
    \caption{Capacitance-voltage (CV) measurements of the selenium thin-film solar cell , where the diode capacitance is analyzed using the diode depletion approximation at various frequencies. Note that the plateau of the apparent carrier concentration is largely independent of frequency suggesting that the capacitance response is reflecting the carrier concentration without interference from transient or dynamic effects.}
    \label{fig:ESI_Figure4}
\end{figure*}

\clearpage

\section{Drive-level capacitance profiling (DLCP)}

For the analysis of drive-level capacitance profiling (DLCP) measurements, the diode capacitance as a function of the AC level ($V_\text{peak}$) was fitted to a second-order polynomial to retrieve the coefficients $C_0$, $C_1$, and $C_2$. The R-squared value was greater than $R^2>0.995$ at all nominal DC voltage biases.

\begin{equation*}
    C = \frac{d\tilde{Q}}{d\tilde{V}} =C_0+C_1 \tilde{V} + C_2 \tilde{V}^2+...
\end{equation*}

\begin{figure*}[h!]
    \centering
    \includegraphics[width=0.48\textwidth,trim={0 0 0 0},clip]{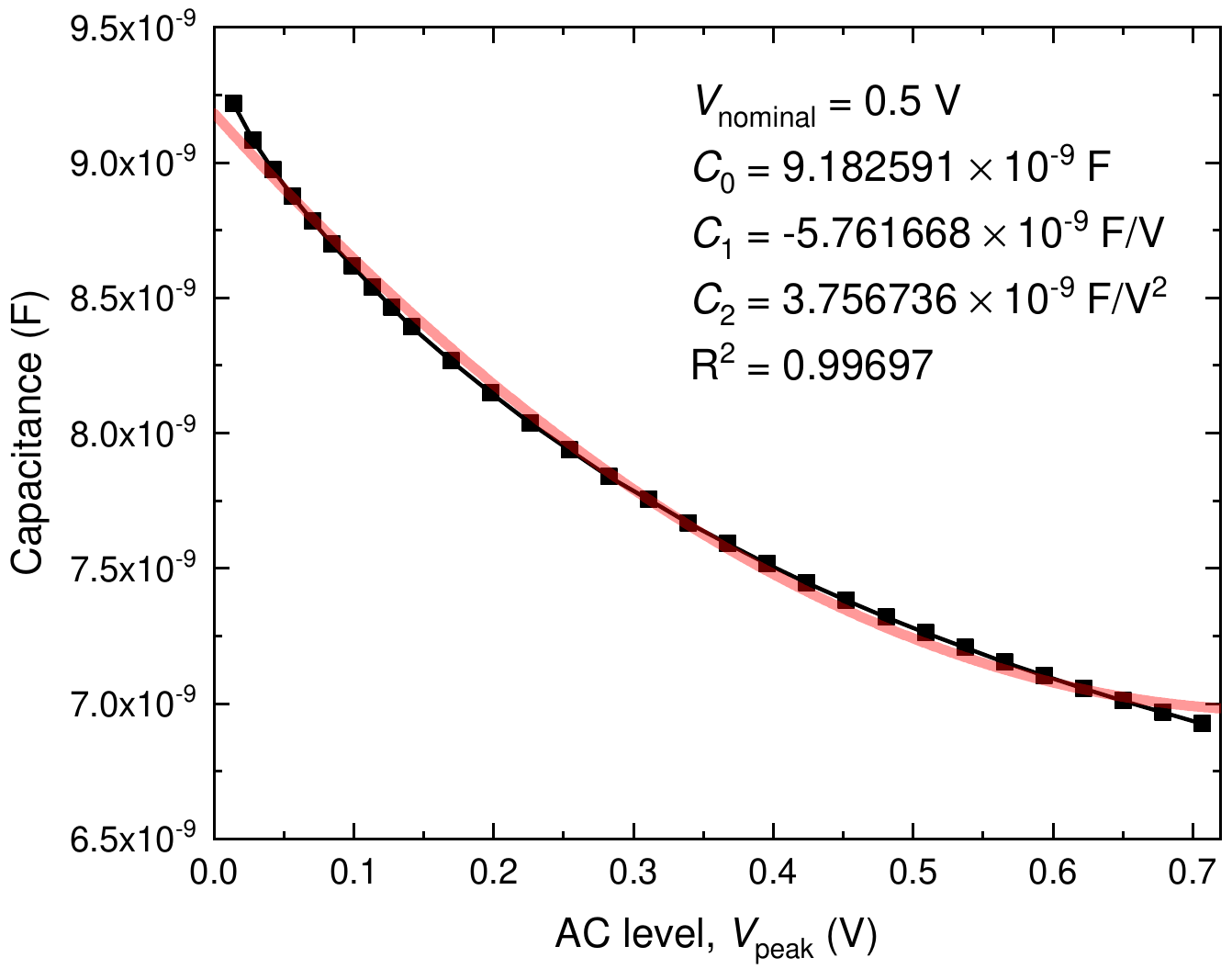}
    \caption{Junction capacitance as a function of AC level fitted using a second-order polynomial.}
    \label{fig:ESI_Figure6}
\end{figure*}

Using the higher order correction coefficients to the junction capacitance, we calculate $N_\text{DLCP}$,

\begin{equation*}
    N_\text{DLCP} = - \frac{C_0^3}{2q_\mathrm{e}\epsilon_0 \epsilon_r A^2 C_1}
\end{equation*}

\noindent where $\epsilon_0$ is the vacuum permittivity, $\epsilon_r$ is the relative permittivity, and $A$ is the active area of the solar cell.\\

We utilized the fact that $N_\text{DLCP}$ is independent of the distance from the barrier interface, and therefore also independent of surface states or near-interface anomalies. This implies that the difference between CV, which captures only the zero-order component of the junction capacitance, and DLCP, which exploits the first-order correction, must arise from the contribution of interfaces or surfaces \cite{michelson1985a}. These contributions are indicated by the frequency or voltage dependence of the first-order term's coefficient.

\clearpage

\section{Depletion approximation}

\begin{figure*}[h!]
    \centering
    \includegraphics[width=\textwidth,trim={0 0 0 0},clip]{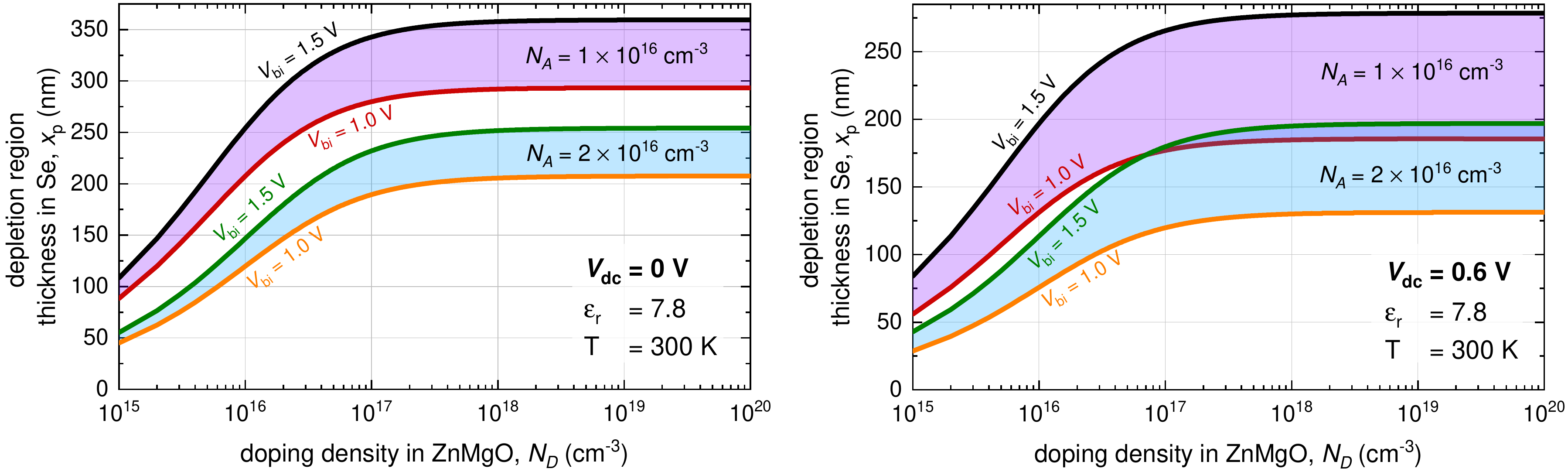}
    \caption{The depletion region thickness in the selenium thin-film solar cell as a function of the doping density in ZnMgO at short-circuit (left) and close to maximum power operating point (right). The built-in voltage is varied between the two extremes derived from the Mott-Schottky plots in Figure S. 5., and the acceptor concentration in selenium is varies between $N_A=1 \times 10^{16}$ cm$^{-3}$ and $N_A=2 \times 10^{16}$ cm$^{-3}$. This further underscores the conclusion from the voltage-biased EQE-spectra, that the acceptor density cannot be on the order of $10^{12}$ cm$^{-3}$, and therefore $p_0\neq N_A$.}
    \label{fig:ESI_Figure3}
\end{figure*}

\clearpage

\section{Injection-level-dependent open-circuit voltage measurements / suns-$V_\text{oc}$}

\begin{figure*}[h!]
    \centering
    \includegraphics[width=\textwidth,trim={0 0 0 0},clip]{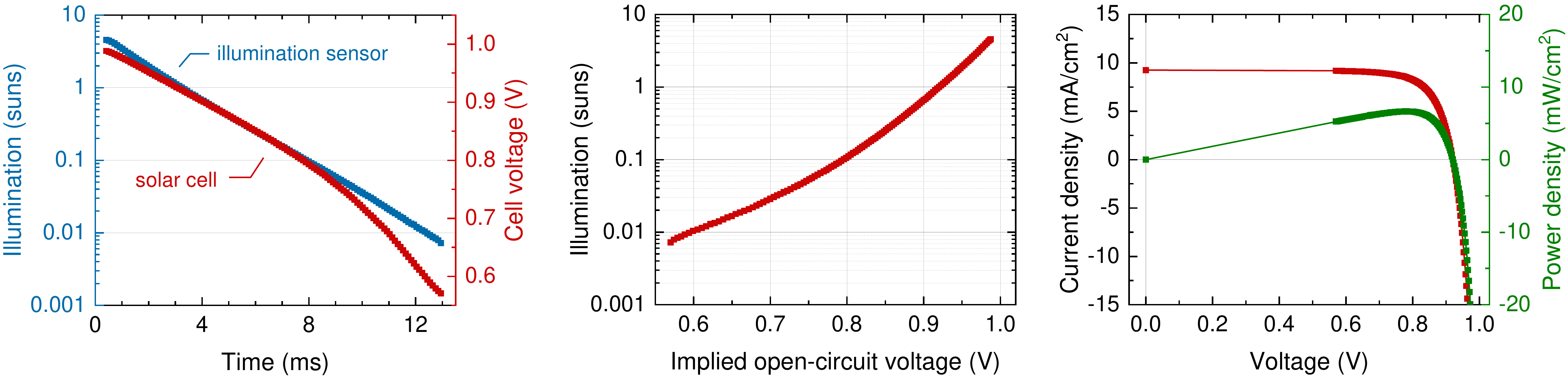}
    \caption{The raw data of the injection-level dependent open-circuit voltage measurements, or suns-$V_\text{oc}$, measured using a white light flasher, and used to reconstruct the current density-voltage and power density-voltage curves of the selenium thin-film solar cell without influence from parasitic transport losses.}
    \label{fig:ESI_Figure5}
\end{figure*}

\clearpage

\section{Raw magnetoresistance data}

\begin{figure*}[h!]
    \centering
    \includegraphics[width=\textwidth,trim={0 0 0 0},clip]{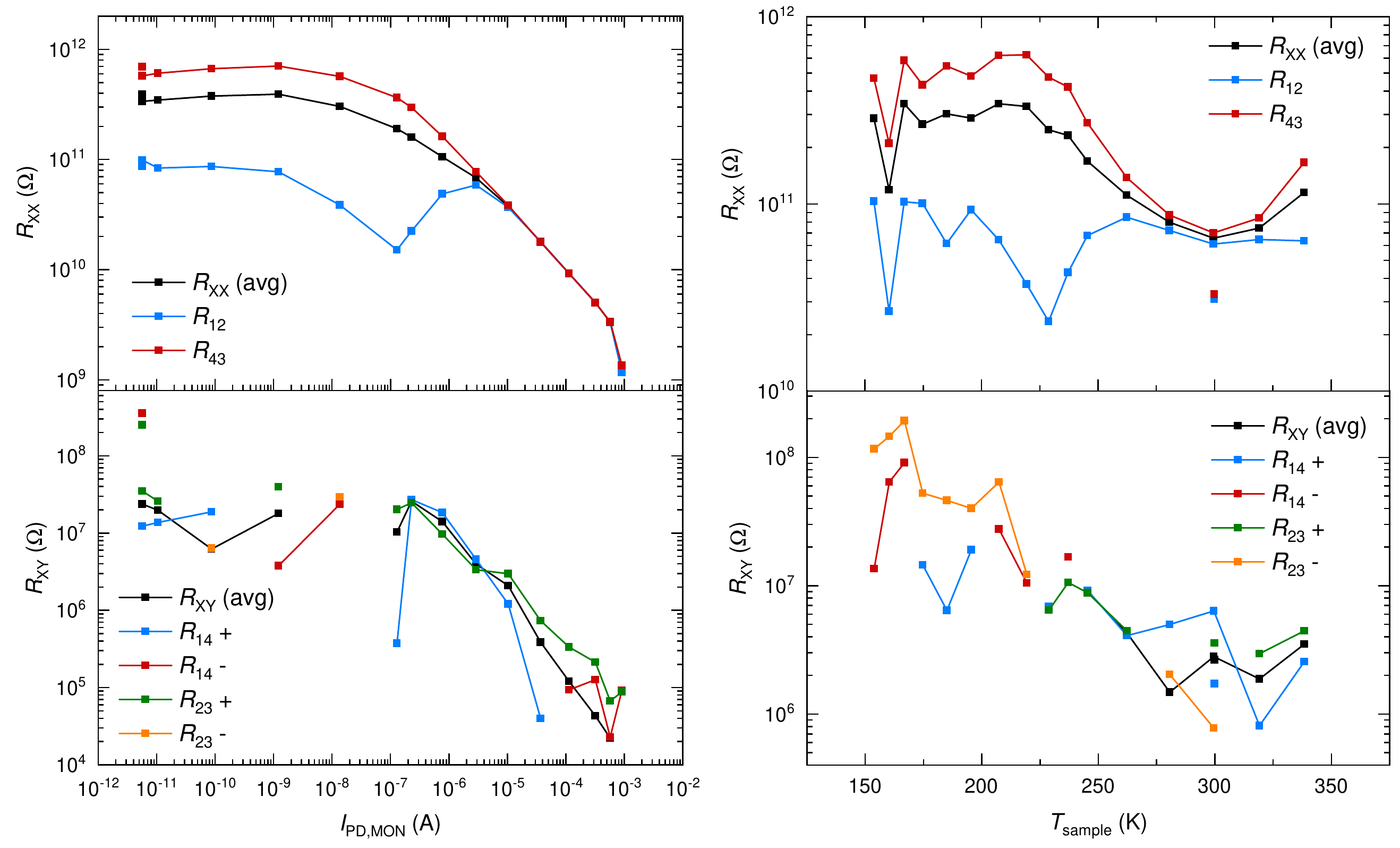}
    \caption{Magnetoresistances ($R_\text{XX}$ longitudinal, $R_\text{XY}$ transverse) as a function of the current readout on the monitoring photodiode and sample temperature, respectively. The subscripts denote the contact terminals, with the current always being sourced from contact 5 to 6, as indicated in the Hall bar schematic in Figure 1 of the main text. The sign of the longitudinal magnetoresistances indicate the sign of the readout, which due to high resistances and low mobilities is easily flipped by noise. However, even under dark conditions, the positive readout prevails indicating a p-type doping in selenium.}
    \label{fig:ESI_Figure1}
\end{figure*}

\clearpage

\section{Temperature control}

\begin{figure*}[h!]
    \centering
    \includegraphics[width=0.5\textwidth,trim={0 0 0 0},clip]{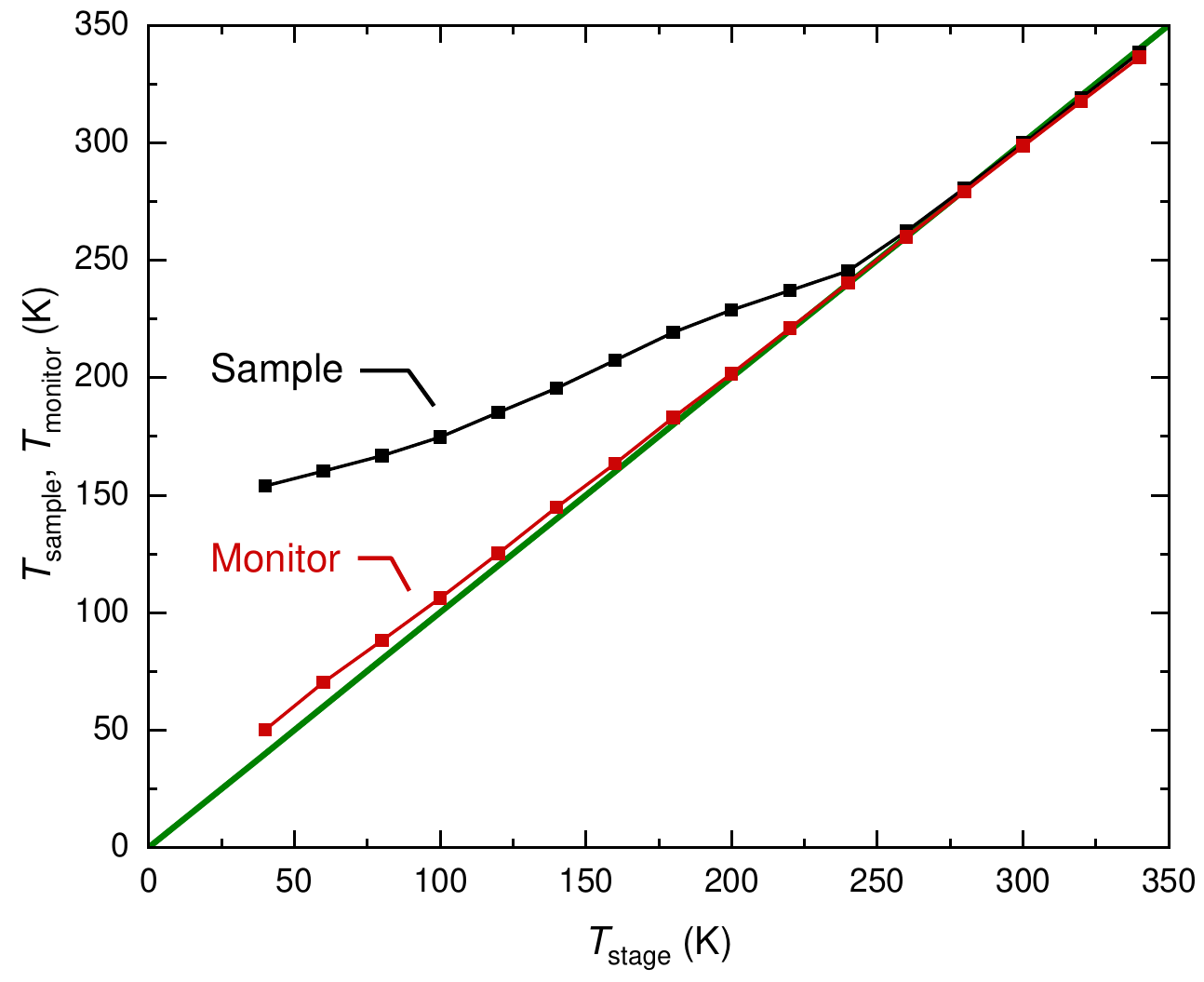}
    \caption{The temperature of the cryostat monitor and the sample as a function of the stage temperature. The sample temperature is measured using a K-type thermocouple mounted on the sample immediately next to the selenium Hall bar.}
    \label{fig:ESI_Figure2}
\end{figure*}

\clearpage


\bibliography{references}